\documentclass[aps,pra, twocolumn,groupedaddress]{revtex4-2}

\usepackage{amsmath}
\usepackage{amssymb}
\usepackage{physics}
\usepackage{graphicx}
\usepackage[colorlinks = true, linkcolor = blue, urlcolor  = blue, citecolor = blue, anchorcolor = blue]{hyperref}
\usepackage{tikz}
\usepackage{comment}
\usepackage[mathscr]{euscript}
\usepackage{subfig}
\usepackage{soul}

\begin{document}
	

\title{Conditions for separability in multiqubit systems with an accelerating qubit using a conditional entropy}


\author{Harsha Miriam Reji}
\thanks{}
\affiliation{Department of Physics, Indian Institute of Technology Dharwad, Dharwad, Karnataka, India - 580011}

\author{Hemant S. Hegde}
\thanks{}
\affiliation{Department of Physics, Indian Institute of Technology Dharwad, Dharwad, Karnataka, India - 580011}

\author{R. Prabhu}
\thanks{}
\affiliation{Department of Physics, Indian Institute of Technology Dharwad, Dharwad, Karnataka, India - 580011}



\begin{abstract}
We study the separability in multiqubit pure and mixed Greenberger–Horne–Zeilinger (GHZ) and W states with an accelerating qubit using the Abe-Rajagopal (AR) $ q $-conditional entropy. We observe that the pure multiqubit GHZ and W states in the inertial : non-inertial bipartition with one of their qubits accelerated will remain non-separable irrespective of the qubit's acceleration. In these systems, we captured the variation of their non-separability with respect to the acceleration of the qubit and the AR $ q $-conditional entropy parameter $ q $. However, in the corresponding multiqubit mixed states obtained by introducing a global noise to the above pure states, one could get stronger conditions on their separability in the inertial : non-inertial bipartition, in terms of the acceleration of the qubit, noise parameter, and the number of qubits in the system, in the asymptotic limit of parameter $ q $. These conditions obtained from the AR $ q $-conditional entropy serve as necessary conditions for separability in such multiqubit states with a relativistic qubit.
\end{abstract}


\maketitle

\section{Introduction}\label{sec1}

A vast majority of quantum information tasks, such as quantum teleportation \cite{PhysRevLett.70.1895, pirandola2015advances,hu2023progress}, quantum dense coding \cite{PhysRevLett.69.2881}, quantum key distribution, \cite{bennett1984proc,PhysRevLett.85.441} etc., rely on quantum correlations present in the quantum systems \cite{RevModPhys.81.865,RevModPhys.84.1655}. Recently, it has been established that the quantum correlations present in quantum systems vary when their subsystems are relativistically accelerated \cite{PhysRevLett.95.120404, PhysRevA.82.042332,PhysRevA.80.052304}. The characterization of quantum correlations and quantum information protocols in quantum systems with accelerated subsystems has led to the emergence of relativistic quantum information. Currently, relativistic quantum information involves the studies that characterize and quantify entanglement in bosonic  \cite{PhysRevLett.95.120404,PhysRevA.81.052305,PhysRevA.92.022334,PhysRevA.81.032320,PhysRevD.86.025026,PhysRevA.84.022303} and fermionic \cite{PhysRevA.79.064301,PhysRevA.83.052306,PhysRevA.98.022320,PhysRevA.80.012314,khan2014non,ahn2008black,PhysRevA.79.042333,PhysRevA.74.032326,PhysRevA.83.022314,khan2014tripartite,PhysRevD.108.125009} modes when their subsystems are in non-inertial frames, entanglement near black holes \cite{PhysRevA.80.042318,PhysRevD.82.064028, ge2008quantum,PhysRevD.78.065015,PhysRevD.105.065007,PhysRevD.82.064006}, entanglement in expanding universe \cite{ball2006entanglement,PhysRevD.82.045030,PhysRevD.79.044027,wu2022quantum, liu2016quantum}, relativistic quantum metrology \cite{tian2015relativistic,ahmadi2014relativistic,PhysRevD.89.065028,wang2014quantum}, quantum discord in relativistic states \cite{PhysRevA.81.052120,PhysRevA.80.052304}, relativistic quantum teleportation \cite{PhysRevLett.91.180404,PhysRevA.106.032432,PhysRevD.95.105005},  relativistic quantum speed limit \cite{khan2015relativistic,haseli2019quantum,khan2022quantum}, etc.

There has been considerable interest in using entropic measures to characterize separability in composite quantum systems in the inertial frame \cite{PhysRevA.70.022316,PhysRevA.70.012102,PhysRevLett.79.5194,Horodecki1994Quantum,PhysRevA.54.1838,Horodecki1996realism}. For pure states, the separability criteria using von Neumann conditional entropy suggest that non-separable quantum systems are more disordered locally than globally \cite{PhysRevLett.86.5184}. In the case of mixed states, the generalized entropic measures have been introduced to better understand and explore the valuable quantum information properties \cite{PhysRevA.63.042104, PhysRevA.65.052323, rossignoli2011generalized, batle2002conditional, batle2003some}. The two prominent families of these generalized entropies are Tsallis  $q$-entropy \cite{tsallis1988possible,tsallis1998role} and $\alpha$-Renyi entropy \cite{RevModPhys.50.221}, which reduce to the von Neumann entropy as $q$ approaches 1. Their positive values help distinguish between global and local disorder in mixed states.

To study the separability of quantum states with an accelerating qubit, we use a generalized form of conditional entropy called the Abe-Rajagopal (AR) $ q $-conditional entropy, derived from the Tsallis $q$-entropy \cite{abe2001nonadditive, abe2002towards}. The AR $ q $-conditional entropy depends on the global versus local spectra of the composite quantum system, and it assumes a negative value for non-separable states. However, 
AR $q$-conditional entropic characterization is not always a necessary and sufficient condition for separability. Notwithstanding this, it can offer stricter limitations on separability compared to traditional methods like the von Neumann conditional entropy  \cite{PhysRevA.76.042337}. Further, a negative AR $q$-conditional entropy indicates the composite quantum state is distillable, as it implies a violation of the reduction criterion \cite{vollbrecht2002conditional}.
Previously, this conditional entropy has been applied to study the separability in a single parameter family of mixed multiqubit states \cite{PhysRevA.76.042337} and Gaussian states \cite{PhysRevA.81.024303}. Here, we use the AR $q$-conditional entropy to characterize the non-separability of several multipartite quantum states in the inertial : non-inertial bipartition.

We initially consider a generalized pure two-qubit Greenberger–Horne–Zeilinger (GHZ) state with one of its qubits under acceleration and characterize its non-separability in the inertial : non-inertial bipartition with respect to system parameters, acceleration of the qubit, and parameter $q$. Since the characterization of non-separability using the AR $q$-conditional entropy uses all the eigenvalues of the state and corresponding subsystem, we introduce an {\it eigenvalue truncation procedure} to handle the infinite eigenvalues in such systems numerically. Subsequently, we extend our non-separability characterization to the pure multiqubit GHZ and W states. Later, we explore the non-separability in mixed states generated by mixing a global noise to the above pure multiqubit GHZ and W states. Since these accelerated mixed states and their subsystems do not possess a block structure to obtain their eigenvalues analytically, we introduce a {\it density matrix truncation procedure} to evaluate the AR $q$-conditional entropy numerically. We derive conditions for separability in these mixed multiqubit states, which depend on their mixing parameter, the acceleration of the qubit, and the number of qubits in the multiqubit state in the asymptotic limit of parameter $q$. The findings on the non-separability conditions of the mixed multiqubit states reveal significant insights when compared with the entanglement measure of logarithmic negativity. The AR $q$-conditional entropy imposes a more stringent condition on non-separability than logarithmic negativity in single qubit accelerated multiqubit mixed states. However, both measures are equivalent for single qubit accelerated pure multiqubit states, a crucial observation in the study of non-separability measures.


This paper is structured as follows: In Sec.~\ref{sec:RQS}, we describe the bosonic quantum system in a non-inertial frame. In Sec.~\ref{sec:AR}, we succinctly describe the quantification of non-separability using the AR $q$-conditional entropy in an arbitrary bipartite quantum system. In Sec.~\ref{sec:pure}, the non-separability of pure multiqubit  GHZ and W states with one of its qubits being accelerated using the AR $q$-conditional entropy is characterized with respect to acceleration, parameter $q$, and number of qubits in the system. In Sec.~\ref{sec:mixed}, we give a detailed account of obtaining the non-separability of mixed multiqubit states obtained by adding a global noise to the pure states considered in Sec.~\ref{sec:pure}. Also, the strongest condition on separability for these states is identified in the asymptotic limit of $q$. A comparison of separability criteria obtained from AR $q$-conditional entropy and logarithmic negativity is presented in Sec.\ref{sec:LN}. In Sec.~\ref{sec:conclusion}, the concluding remarks are presented.

\begin{figure}
	\centering \includegraphics[width=0.95\linewidth]{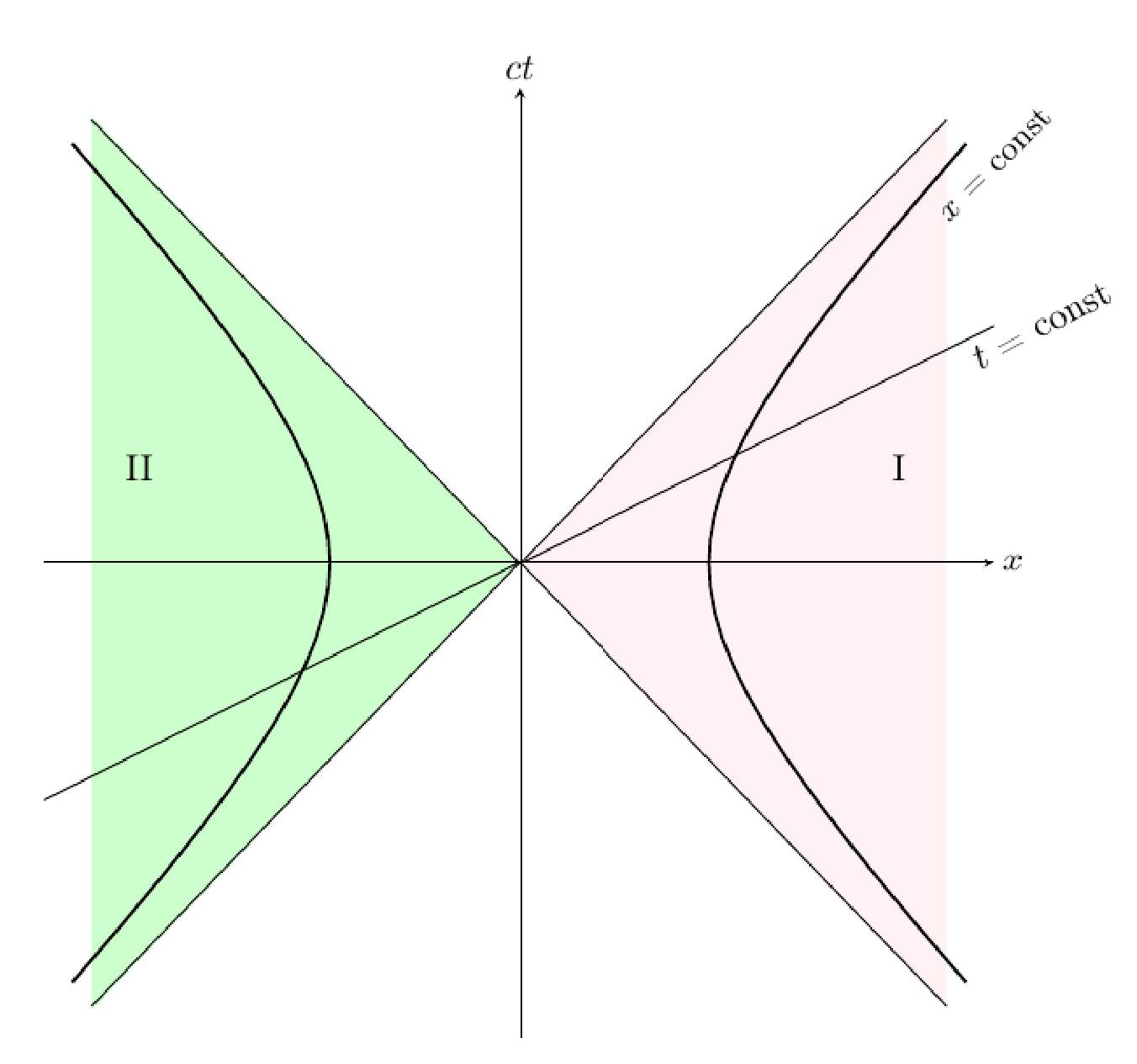}
	\caption{Schematic representation of Rindler space-time \cite{PhysRevLett.91.180404}, where $ t $ is the proper time, and $ x =$ const is the trajectory of a qubit with constant acceleration. The pink-shaded region to the right of the vertical axis denotes region I and the green-shaded region to the left of the vertical axis denotes region II of Rindler space-time.}
	\label{fig:Rindler}
\end{figure}

\section{Relativistically Accelerated Quantum State}\label{sec:RQS}

Consider an arbitrary $2$-qubit state $\rho_{AB}$ in an inertial frame shared between the two observers, Alice and Bob. When Bob undergoes a relativistic acceleration, henceforth called Rob (relativistically accelerated Bob), his qubit experiences a different spacetime around it. While Minkowski coordinates are commonly employed to depict the qubits in the inertial frames, they are unsuitable for describing the motion of an accelerated qubit. In non-inertial frames, Rindler coordinates are more appropriate for describing the accelerating qubits. The schematic representation of the Rindler spacetime is provided in Fig.~\ref{fig:Rindler}. The Rindler spacetime is divided into two distinct regions, region I (pink shade) and region II (green shade), containing all points mapped from the Minkowski coordinates. Using Bogoliubov transformations, the vacuum state and the single excited state in Minkowski coordinates will individually take the form of a two-mode squeezed state in Rindler spacetime \cite{PhysRevLett.95.120404,PhysRevLett.91.180404}, i.e.,
\begin{equation} \label{eq:R0}
	\ket{0_k}_M  = \frac{1}{\cosh r} \sum_{n=0}^{\infty} \tanh^n r \ket{n_k}_{I} \otimes \ket{n_k}_{II}, 
\end{equation}
and 
\begin{equation} \label{eq:R1}
	\ket{1_k}_M = \frac{1}{\cosh^2 r} \sum_{n=0}^{\infty} \tanh^n r \sqrt{n+1}\ket{(n+1)_k}_{I} \otimes \ket{n_k}_{II}
\end{equation}
respectively. Here $\ket{n_k}_{I}  $ and $ \ket{n_k}_{II} $ are the number states of  $ k^{\text{th}}$ mode in the region I and region II of the Rindler spacetime respectively. The RHS of Eqs.~(\ref{eq:R0}) and (\ref{eq:R1}) corresponds to a two-mode squeezed states, with the squeezing parameter $ r $ being related to the acceleration $ a $ of Rob's qubit as $	\cosh r = ( 1 - e^{-2 \pi \abs{k} c/a})^{-1/2} $. Here, $ c $ is the speed of light in a vacuum. 

A composite state shared between Alice and Bob in the inertial frame $\rho_{AB}$ transforms as $\rho_{A R_I R_{II}}$ when Bob's qubit is accelerated. However, the qubit moving in one of the regions of the Rindler spacetime cannot travel into the other region as they are causally disconnected. Without any loss of generality, one can trace out the modes of the region II ($R_{II}$) of the accelerating qubit. Since a part of the accelerating state has been traced out, the final state between Alice ($A$) and Rob's modes in the region I ($R_I$) is a {\it mixed} state and is in the basis $ \{ \ket{0}_M\ket{n}_{I}, \ket{0}_M\ket{n+1}_{I}, \ket{1}_M\ket{n}_{I}, \ket{1}_M\ket{n+1}_{I} \} $ with $ n $ going from zero to $ \infty $.  Henceforth, we will drop all the subscripts $ M $ and $ I $ in the inertial basis, $ \{\ket{0}, \ket{1}\}, $ as well as accelerating qubit's basis, $ \{\ket{n}, \ket{n+1}\}, $ and denote the resultant mixed state of Alice and Rob's mode in the region I by $ \rho_{AR} $ and Rob's subsystem as $ \rho_R $.

\section{Abe-Rajagopal $ q $-conditional entropy}\label{sec:AR}

The Tsallis $q$-entropy gives a non-extensive generalization of Shannon entropy  \cite{tsallis1988possible,tsallis1998role}, which is always positive and adheres to the non-additivity relationship for bipartite state $ \rho_{XY} $, i.e.,  $S_q^T(\rho_{XY}) = S_q^T(\rho_{X}) + S_q^T(\rho_{Y}) + (1 - q)\,\, S_q^T(\rho_{X}) \,\, S_q^T(\rho_{Y})$, where $ S_q^T(\rho)   $ is given by
\begin{equation}\label{eq:Tsallis}
	S_q^T(\rho)  = \frac{\Tr [ \rho^q ] - 1}{1-q}.
\end{equation}
Here $\Tr $ denotes the trace of the corresponding density matrix. However, this non-additivity condition is violated by the state $\rho_{XY}$ when subsystems $X$ and $Y$ exhibit long-range interactions, such as entanglement, which persists even when they are spatially separated by long distance. To address this, a conditional form of Tsallis $q$-entropy was introduced by Abe et al. \cite{abe2001nonadditive} called the Abe-Rajagopal $q$-conditional entropy and is defined as
\begin{equation}\label{eq:AR}
	S_q(X|Y) = \frac{S^T_q(\rho_{XY}) - S^T_q(\rho_{Y})} {1 + (1-q)S^T_q(\rho_{Y}) },
\end{equation}
where $ S^T_q(\rho_{XY}) $ and $ S^T_q(\rho_{Y}) $ denote the Tsallis $ q $-entropies  associated with the composite system $ \rho_{XY} $ and its subsystem $ \rho_{Y}$. The computable form of this conditional entropy for any bipartition $ X:Y $ in a multiqubit state can be obtained by plugging in the Eq.~(\ref{eq:Tsallis}) in Eq.~(\ref{eq:AR}) and is given by 
\begin{align}\label{eq:AR1}
	S_q(X|Y) & = \nonumber \frac{1}{q-1} \left[ 1 - \frac{\Tr[\rho_{XY}^q]} {\Tr[\rho_{Y}^q]} \right], \\
	& = \frac{1}{q-1} \left[ 1 - \frac{\sum_n \lambda^q_n(\rho_{XY})}{\sum_m \lambda^q_m(\rho_{Y})}\right],
\end{align}
where $ \lambda_n $ and $ \lambda_m $ are the eigenvalues of the whole system $ \rho_{XY}$ and one of its subsystem $\rho_{Y}$ respectively. This conditional entropy adheres to the non-additivity relation: $S_q^T(\rho_{XY}) = S_q(X|Y) + S_q^T(\rho_Y) + (1 - q)S_q(X|Y)S_q^T(\rho_{Y})$. The AR $ q $-conditional entropy yields negative values for the non-separable states, and hence, the positivity of the AR $ q $-conditional entropy forms a criterion for capturing the separability in bipartite states. From here onwards, we represent $S_q(X|Y)$ and  $\mathscr{S}_q(X|Y)$ as the AR $q$-conditional entropy for pure $\rho_{XY}$ and mixed $\varrho_{XY}$ states with one of their qubit being accelerated respectively.

\section{PURE MULTIQUBIT STATES WITH AN ACCELERATING QUBIT}\label{sec:pure}

In this section, we consider pure multiqubit GHZ and W states with an accelerated qubit and characterize their non-separability in the inertial : non-inertial bipartition using the AR $q$-conditional entropy given in Eq.~(\ref{eq:AR1}) with respect to their system parameters, parameter $q$ and acceleration of the qubit in the non-inertial frame. 
\subsection{Pure multiqubit GHZ states}\label{subsec:evt}

\begin{figure*}[htpb]
	\centering
	\subfloat[]{
		\includegraphics[width=0.45\linewidth]{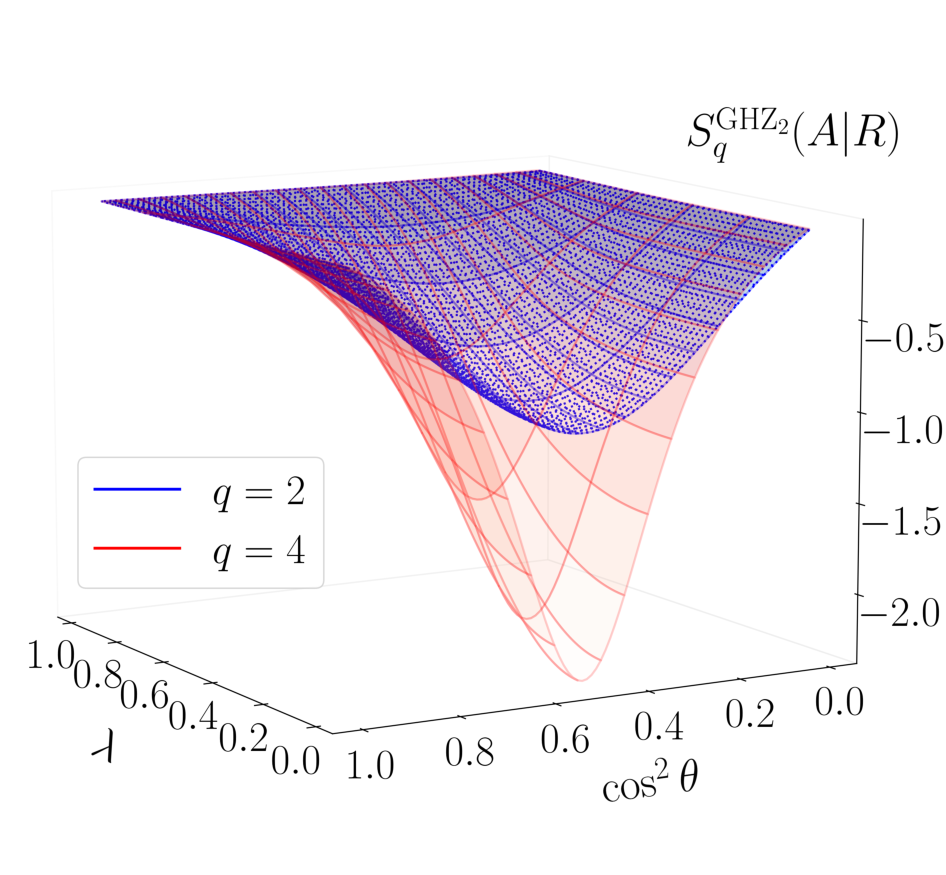}
		\label{fig:CAR_pure2qubit_1in1acc_q2_3D}}
		\hspace{1cm}
	\subfloat[]{
		\includegraphics[width=0.4\linewidth]{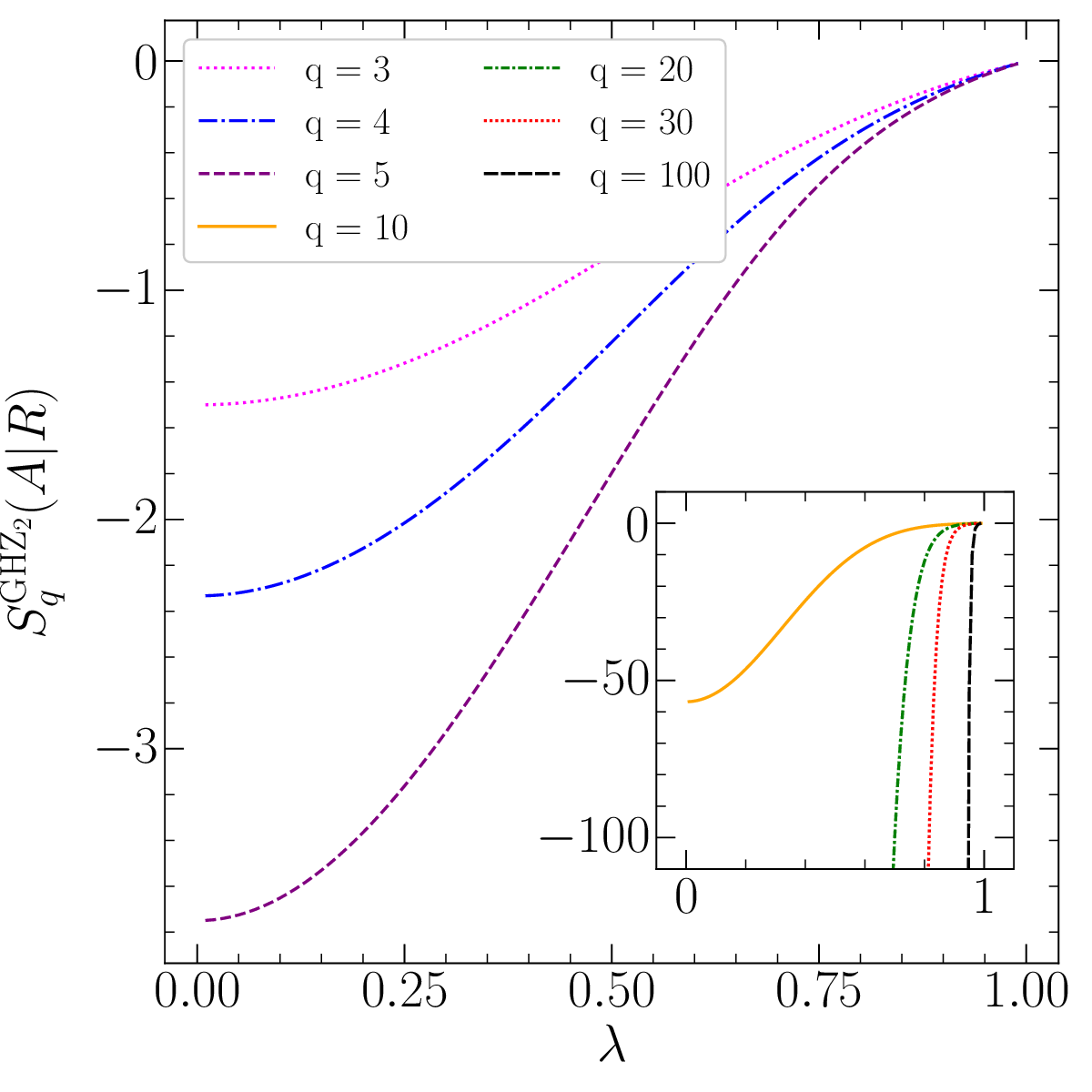}
		\label{fig:CAR_pure2qubit_inset}}
	\caption{(a) The variation of the Abe-Rajagopal $ q $-conditional entropy as a non-separability feature of a single qubit accelerated generalized pure $ 2 $-qubit GHZ state as a function of $ \cos^2 \theta $ and $ \lambda$ for parameter $ q = 2 $ and $ 4 $. The upper blue solid mesh corresponds to $ q=2 $, and the lower red solid mesh corresponds to $ q=4 $.  The $ S^{\text{GHZ}_2}_q(A|R) $ is negative everywhere, implying that the state is non-separable. For this plot, we have chosen 100 values each of $\cos^2 \theta$ and $\lambda$ to create the surface, and the meshes are drawn for representation purposes. (b) The variation of the AR $ q $-conditional entropy of pure $ 2 $-qubit GHZ state with $\theta=\pi/4$ as a function of $ \lambda $ for different values of the parameter $ q $ is depicted in this figure.}
\end{figure*} 

\noindent\textbf{Generalized pure $ 2 $-qubit GHZ state:} A generalized pure $2$-qubit GHZ state shared between Alice and Bob in the inertial frame is given by
\begin{equation}\label{eq:bell}
	\ket{\text{GHZ}_2} =  \cos{\theta}\ket{00} + \sin{\theta}\ket{11}; \; \cos^2{\theta}+\sin^2{\theta} =1,
\end{equation}
which belongs to the Hilbert space ${H}_A \otimes {H}_B$. Let Bob's qubit be relativistically accelerated with respect to Alice's qubit, and the resulting composite state, $ \rho^{\text{GHZ}_2}_{A R_I R_{II}} $, is obtained by using the transformations of $\ket{0}$ and $\ket{1}$ from Minkowski spacetime to Rindler spacetime as given in Eqs.~(\ref{eq:R0}) and (\ref{eq:R1}). We now obtain the marginal mixed density matrix between Alice and Rob's modes in region I, by following the recipe given in Sec. \ref{sec:RQS}, as
\begin{align}\label{eq:2qubit_AR_ghz}
	&\rho_{AR}^{\text{GHZ}_2} =\nonumber \frac{1}{\cosh^{2}{r}} \sum_{n} \tanh^{2n} {r}  \\
	& \hspace{0.5cm}\nonumber \times \Big[  \cos^2{\theta} \ket{0n}\bra{0n} +  \frac{n+1}{\cosh^{2}{r}} \sin^2{\theta} \ket{1n+1}\bra{1n+1} \\
	& \hspace{0.5cm}+ \frac{\sqrt{n+1}}{\cosh{r}} \cos{\theta} \sin{\theta} \,\,\big(\ket{1n+1}\bra{0n} +  \ket{0n} \bra{1n+1}\big) \Big] .	
\end{align}
The structure of this density matrix contains $ 2\times 2 $ blocks along the diagonal, with the remaining off-diagonal elements being zero and is given by 
\begin{equation}\label{eq:block_w}
	\rho_{AR}^{\text{GHZ}_2} = \frac{1}{\cosh^{2}{r}}\begin{pmatrix}	
		0 & & & & &\\
		& \Delta_0 &&&&\\
		&& \Delta_1 &&&\\
		&&& \ddots && \\
		&&&& \Delta_n & \\
		&&&&& \ddots
	\end{pmatrix},
\end{equation}
where
\begin{equation}
	\Delta_n(\rho_{AR}^{\text{GHZ}_2}) = \tanh^{2n} {r} \begin{pmatrix}	
		\cos^2{\theta} & \alpha_n \cos{\theta} \sin{\theta}  \\
		\alpha_n \cos{\theta}  \sin{\theta}  &  \alpha_{n}^2 \sin^2{\theta}  \\
	\end{pmatrix}
\end{equation}
with $ \alpha_{n} = \frac{\sqrt{n+1}}{\cosh r} $.
The eigenvalues of this $ n^{\text{th}} $ block of the state $ \rho^{\text{GHZ}_2}_{A R} $ are $0$ and 
\begin{equation}\label{eq:pure_2qubitghz_AR_EV}
	\Lambda_n = \frac{ \tanh^{2n} r}{\cosh^2 r} \left( \cos^2\theta + \frac{n+1}{\cosh^2 r} \sin^2 \theta\right).
\end{equation} 
Therefore, $ \Tr(\rho^{\text{GHZ}_2}_{A R})=\sum_{n=0}^{\infty} \Lambda_n = 1 $. 

The marginal density matrix  $\rho^{\text{GHZ}_2}_{R}$ corresponding to the modes of Rob in the region I of Rindler spacetime is computed by tracing out  Alice's subsystem $A$, i.e.,  $ \rho^{\text{GHZ}_2}_R = \Tr_A (\rho^{\text{GHZ}_2}_{AR}) $ and is given by 
\begin{align}\label{eq:ghz_pure_R}
	\rho_{R}^{\text{GHZ}_2} \nonumber = &\frac{1}{\cosh^2 r} \sum_{m=0}^{\infty} \tanh^{2m} r \Big[\cos^2 \theta  \ket{m}\bra{m} \\
	&+ \frac{ m+1 }{\cosh^2 r} \sin^2 \theta   \ket{m+1}\bra{m+1} \Big].
\end{align} 
This is an infinite dimensional diagonal matrix whose $ m^{\text{th}} $  eigenvalue is given by 
\begin{equation}\label{eq:pure_2qubitghz_R_EV}
	\Lambda_m = \frac{ \tanh^{2m} r}{\cosh^2 r} \left( \cos^2\theta + \frac{m}{\sinh^2 r} \sin^2 \theta\right).
\end{equation}
Here, $ \Tr(\rho^{\text{GHZ}_2}_{R})=\sum_{m=0}^{\infty} \Lambda_m = 1 $. 

To characterize the non-separability  of the state $ \rho^{\text{GHZ}_2}_{AR} $ given in Eq.~(\ref{eq:2qubit_AR_ghz}), we substitute the eigenvalues $ \Lambda_n $'s of $ \rho^{\text{GHZ}_2}_{AR} $ (see Eq.~(\ref{eq:pure_2qubitghz_AR_EV})) and $ \Lambda_m $'s of $ \rho^{\text{GHZ}_2}_R $ (see Eq.~(\ref{eq:pure_2qubitghz_R_EV})) in the AR $q$-conditional entropy given in  Eq.~(\ref{eq:AR1}) to obtain 
\begin{align}\label{eq:CAR_pure_2qubit_ghz}
	&S^{\text{GHZ}_2}_q(A|R)  \nonumber =\frac{1}{q-1} \\
	& \times \left[ 1 - {\displaystyle \sum_{n=0}^{\infty} \left( \tanh^{2n} r \left(\cos^2{\theta} + \sin^2{\theta} \frac{n+1}{\cosh^2 r}\right)  \right)^q  \over \displaystyle \sum_{m=0}^{\infty} \left( \tanh^{2m} r \left(\cos^2{\theta} + \sin^2{\theta} \frac{m}{\sinh^2 r}\right)\right)^q} \right].
\end{align}
The negative values of this AR $ q $-conditional entropy $S^{\text{GHZ}_2}_q(A|R)$ characterize the non-separability between Alice's qubit and Rob's mode in the region I of the Rindler spacetime. As the above expression contains summations running up to infinity, we have to evaluate $S^{\text{GHZ}_2}_q(A|R)$ numerically to study its behaviors with respect to the state parameter, acceleration parameter $r$, and parameter $q$.

\noindent {\it Eigenvalue truncation procedure}: To numerically compute this conditional entropy, we make use of the fact that the sum of the non-zero eigenvalues of $\rho^{\text{GHZ}_2}_{A R}$ and $\rho^{\text{GHZ}_2}_{R}$ tends to $ 1$ as their respective number of eigenvalues $ n $ and $ m $ approaches infinity. We achieve this numerically by fixing a large value of $ n\simeq k $ and $ m\simeq k $ such that the sum of $ k $ eigenvalues approaches $ 1 $. By inspection, we found that this is satisfied for both the states $\rho^{\text{GHZ}_2}_{A R}$ and $\rho^{\text{GHZ}_2}_{R}$ when $ k\approx 10^4 $. Now, using these $k$ number of eigenvalues of $\rho^{\text{GHZ}_2}_{A R}$ and $\rho^{\text{GHZ}_2}_{R}$, we numerically compute the AR $ q $-conditional entropy given in Eq.~(\ref{eq:CAR_pure_2qubit_ghz}) and characterize the non-separability of the state in Eq.~(\ref{eq:2qubit_AR_ghz}) with respect to the parameter $ q $, the state parameter $\theta$, and Rob's acceleration $ r $.

We now introduce the modified acceleration parameter $\lambda = \tanh r$, which carries all the characteristics of Rob's acceleration $ r $, i.e., we use  $\lambda \in (0,1)$ instead of $r \in (0,\infty)$. For simplicity, we call $\lambda$ as the Rob's acceleration. The AR $ q $-conditional entropy for the state given in Eq.~(\ref{eq:2qubit_AR_ghz}) is then plotted as a function of its state parameter $ \theta $ and acceleration of Rob $\lambda$ in Fig.~\ref{fig:CAR_pure2qubit_1in1acc_q2_3D} for $ q=2$ (upper blue surface) and $4 $ (lower red surface). We observe that the $ S^{\text{GHZ}_2}_q(A|R) $ always remains less than zero, implying that the state remains non-separable for any values of $\lambda$ and $\theta$. A higher negative value of the AR $q$-conditional entropy implies higher non-separability; however, this does not necessarily imply a higher entanglement in mixed states. Note that, $ S^{\text{GHZ}_2}_q(A|R) $ attains maximum non-separability at $ \theta = \pi/4 $, irrespective of the parameter $ q $. When $ \theta = \pi/4$, Eq.~(\ref{eq:bell}) corresponds to a maximally entangled $ 2 $-qubit Bell state in an inertial frame. Hence, in Fig.~\ref{fig:CAR_pure2qubit_inset}, we plot the characterization of $ S^{\text{GHZ}_2}_q(A|R) $ for the state $\rho^{\text{GHZ}_2}_{A R}$, given in Eq.~(\ref{eq:2qubit_AR_ghz}), with the initial state chosen as the Bell state, with respect to $\lambda$ for various values of $ q $. We observe that irrespective of the value of $ q $, the non-separability of $\rho^{\text{GHZ}_2}_{A R}$ will be high at lower values of acceleration and decreases with the increase in acceleration, in particular, $ S^{\text{GHZ}_2}_q(A|R) $ tends to $ 0 $ as $ \lambda$ approaches  $ 1 $. As the value of $ q $ increases, the AR $q$-conditional entropy suggests that the state $\rho^{\text{GHZ}_2}_{AR}$ may transition towards separability in the limit $\lambda$ approaches  $1$. However,  $ S^{\text{GHZ}_2}_q(A|R) $ approaching $ 0 $ becomes more and more steeper at large values of both acceleration $\lambda$ and parameter $ q $.

\noindent\textbf{Generalized pure $ 3 $-qubit GHZ state:}
Let us consider a generalized pure $3$-qubit GHZ state given by 
\begin{equation}\label{eq:gGHZ}
	\ket{\text{GHZ}_3} = \cos{\theta} \ket{000} +  \sin{\theta} \ket{111}; \; \cos^2{\theta}+\sin^2{\theta} =1,
\end{equation} 
shared between Alice and Bob, with the first two of its qubits in Alice's possession while the remaining qubit is in Bob's possession, i.e., they belong to the $H_{A_1} \otimes H_{A_2} \otimes H_B$ Hilbert space. By accelerating Bob's qubit, we study the non-separability between inertial Alice's two qubits and modes of Rob in the region I of Rindler spacetime by exploring the AR $q$-conditional entropy. Using the recipe given in Sec. \ref{sec:RQS}, we get the composite state of the inertial $ A_1A_2 $ and non-inertial $ R $ as $\rho_{A_1 A_2 R_I R_{II}}^{\text{GHZ}_3}$, and the state $ \rho_{A_1 A_2 R}^{\text{GHZ}_3} $ as the mixed subsystem after tracing out the modes of Rob's qubit in region II. The density matrix of $\rho_{A_1A_2R}^{\text{GHZ}_3}$ is given by
\begin{align}\label{eq:dm_pure_ghz3}
	&\rho_{A_1A_2R}^{\text{GHZ}_3} \nonumber = \frac{1}{\cosh^2 r} \sum_{n=0}^{\infty} \tanh^{2n} r \Big[\cos^2{\theta} \ket{00n}\bra{00n} \\
	\nonumber & + \frac{\sqrt{n+1} }{\cosh r} \cos{\theta} \sin{\theta} \Big(\ket{00n}\bra{11n+1} +  \ket{11n+1}\bra{00n} \Big) \\
	&+ \frac{n+1}{\cosh^2 r} \sin^2{\theta}   \ket{11n+1}\bra{11n+1} \Big].
\end{align} 
Generally, there are two independent $ q $-conditional entropies associated with arbitrary $ 3 $-qubit states \cite{PhysRevA.65.052323}, which are given by
\begin{equation} \label{eq:CAR_def_AB|C}
	S_q(AB|C)  = \frac{S^T_q(\rho_{ABC}) - S^T_q(\rho_{C})} {1 + (1-q)S^T_q(\rho_{C}) }
\end{equation}	
and
\begin{equation} 
	S_q(A|BC)  = \frac{S^T_q(\rho_{ABC}) - S^T_q(\rho_{BC})} {1 + (1-q)S^T_q(\rho_{BC}) },
\end{equation} 
where $ S^T_q(\rho_i) $ is the Tsallis $q$-entropy of any state $\rho_i$. Since one of the qubits in the shared initial state is accelerating in a non-inertial frame, a simultaneous measurement cannot be performed on the inertial and the non-inertial qubits together. Hence, we choose the conditional entropy such that the inertial Alice's qubits $ A_1,A_2 $ are conditioned on the non-inertial Rob's modes $ R $. Therefore, we proceed to investigate the separability using the AR $ q $-conditional entropy $  S^{\text{GHZ}_3}_q(A_1 A_2|R) $ in the $ A_1A_2:R$ bipartition for the state $\rho^{\text{GHZ}_3}_{A_1 A_2 R}$ given in Eq.~(\ref{eq:dm_pure_ghz3}).

To get the eigenvalues of the states $\rho^{\text{GHZ}_3}_{A_1 A_2 R}$  and $\rho^{\text{GHZ}_3}_{R}$ to evaluate  $  S^{\text{GHZ}_3}_q(A_1 A_2|R) $, we make use of the fact that their density matrices have the same non-zero eigenvalues as that of $\rho^{\text{GHZ}_2}_{A R}$ and $\rho^{\text{GHZ}_2}_{ R}$ respectively. Hence, the behavior of  $ S^{\text{GHZ}_3}_q(A_1 A_2|R) $  with respect to state parameter $\theta$, acceleration $\lambda$ and $ q $ is exactly same as the behavior of the AR $ q $-conditional entropy for pure $ 2 $-qubit GHZ state with one of its qubit in acceleration $ S^{\text{GHZ}_2}_q(A|R) $ as given in Fig.~\ref{fig:CAR_pure2qubit_1in1acc_q2_3D} and \ref{fig:CAR_pure2qubit_inset}.
For logistic reasons, which will be clear later, we now study the non-separability of pure $N$-qubit GHZ state instead of the generalized $N$-qubit GHZ state.

\begin{figure}
	\centering	\includegraphics[width=0.95\linewidth]{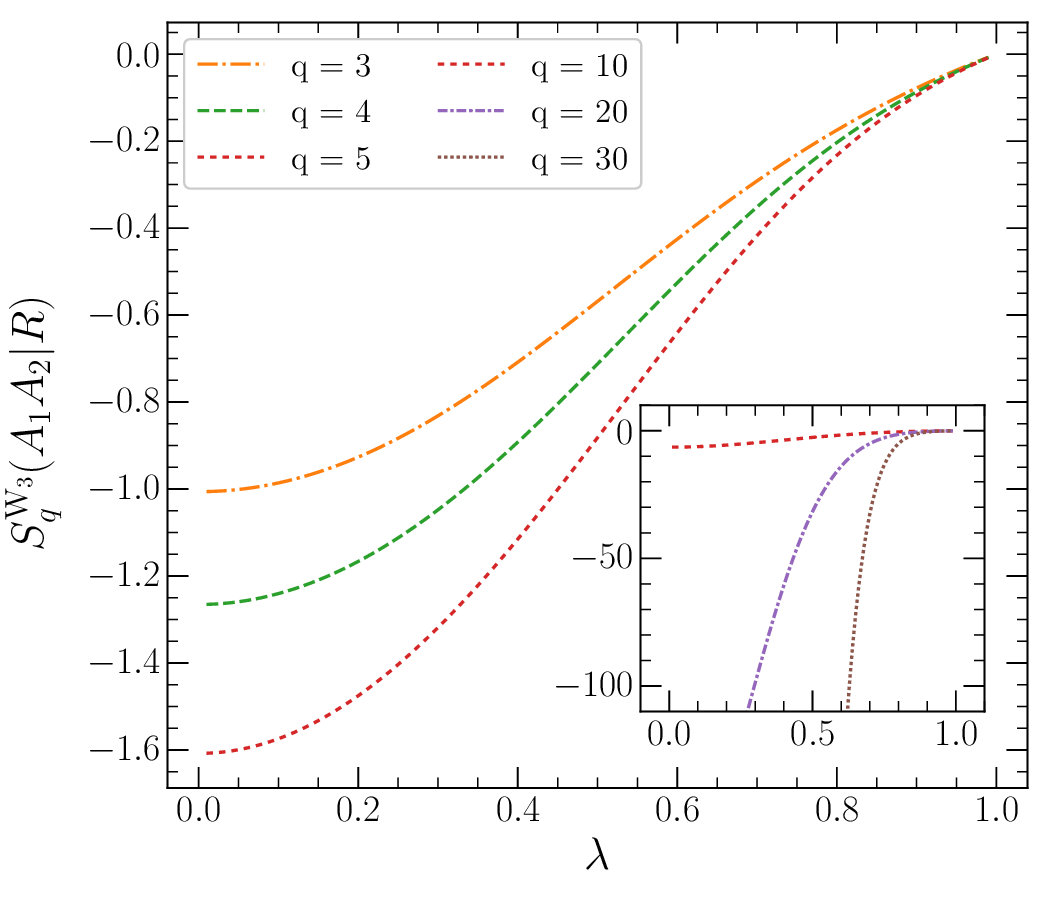}	
	\caption{The variation of the AR $ q $-conditional entropy for pure $ 3 $-qubit W state as a function of acceleration parameter $ \lambda $ for various values of $ q $. We have chosen $\theta=\cos^{-1} (1/\sqrt{3})$ and $\phi=\pi/4$ for this plot.}
	\label{fig:CAR_pure_w_AB_C_inset}
\end{figure} 

\noindent\textbf{Pure $ N $-qubit GHZ state:}
Here, we generalize the characterization of the non-separability of the pure GHZ states to $N$-qubit GHZ state when one of its qubits is accelerated using the AR $q$-conditional entropy. Then, the pure $N$-qubit GHZ state is given by
\begin{equation}
	\ket{\text{GHZ}_N} = \frac{\ket{0}^{\otimes N}+ \ket{1}^{\otimes N}}{\sqrt{2}}, \label{eq:ghz} 
\end{equation} 
belonging to  Hilbert space $H_{A_1} \otimes H_{A_2} \otimes \cdots \otimes H_{A_{N-1}} \otimes  H_B$, i.e., the first $ N-1 $ qubits are with Alice and the last qubit is in Bob's possession. The choice of the qubit, among pure $N$-qubit GHZ states, to be accelerated does not change the study of non-separability among the inertial and non-inertial qubits as the density matrix structure of the composite state and its subsystem remain the same.  When Bob accelerates, the resulting composite state of inertial Alice and non-inertial Rob, denoted as $ \rho_{A_1 A_2  \cdots  A_{N-1} R_I R_{II}}^{\text{GHZ}_N} $, is obtained by using the transformations from Minkowski to Rindler spacetime given in Eqs.~(\ref{eq:R0}) and (\ref{eq:R1}). Then, $ \rho_{A_1 A_2  \cdots  A_{N-1} R}^{\text{GHZ}_N} $ after tracing out one of the disjoint regions in Rindler spacetime (region II) is given by 
\begin{align}\label{eq:NGHZ_AR}
	\nonumber\rho_{A_1 A_2  \cdots  A_{N-1} R}^{\text{GHZ}_N} &=  \frac{1}{2 \cosh^2 r} \sum_{n=0}^{\infty} \tanh^{2n} r \\ 
	& \nonumber \times \Big[ \ket{00 \cdots 0n} \bra{00 \cdots 0n}  \\
	& \nonumber +  \frac{\sqrt{n+1}}{\cosh r} \Big(  \ket{00 \cdots 0n} \bra{11 \cdots 1 n+1}    \\
	& \nonumber + \ket{11 \cdots 1 n+1}\bra{00 \cdots 0n} \Big) \\
	&  + \frac{n+1}{\cosh^2 r} \ket{11 \cdots 1 n+1} \bra{11 \cdots 1 n+1}\Big].
\end{align}
The Rob's subsystem $\rho_R^{\text{GHZ}_N}$ is equivalent to Eq.~(\ref{eq:ghz_pure_R}) with $\theta=\pi/4$. These two density matrices have the same non-zero eigenvalues as that of $\rho_{AR}^{\text{GHZ}_2}$ and $\rho_{R}^{\text{GHZ}_2}$ given in Eqs.~(\ref{eq:pure_2qubitghz_AR_EV}) and (\ref{eq:pure_2qubitghz_R_EV}) with $\theta=\pi/4$.
Therefore, the behavior of the AR $ q $-conditional entropy remains the same with respect to $\lambda$ for various values of $q$ as seen in Fig.~\ref{fig:CAR_pure2qubit_inset}. Increasing the number of qubits in the inertial frame does not affect the non-separability behavior of pure $N$-qubit GHZ state when any of its qubits is accelerated.

Hence, we can conclude that the pure multiqubit GHZ state with one of its qubits under acceleration always remains non-separable throughout the range of $\lambda$ and choices of $q$. The states are more non-separable when the acceleration is low, and the non-separability reduces as the acceleration increases, irrespective of the parameter $ q $ value. The non-separability of the state in Eq.~(\ref{eq:NGHZ_AR}) approaching  $0$ becomes more and more steeper at large values of $\lambda$ and $q$.

\begin{figure}
	\centering
	\includegraphics[scale=0.5]{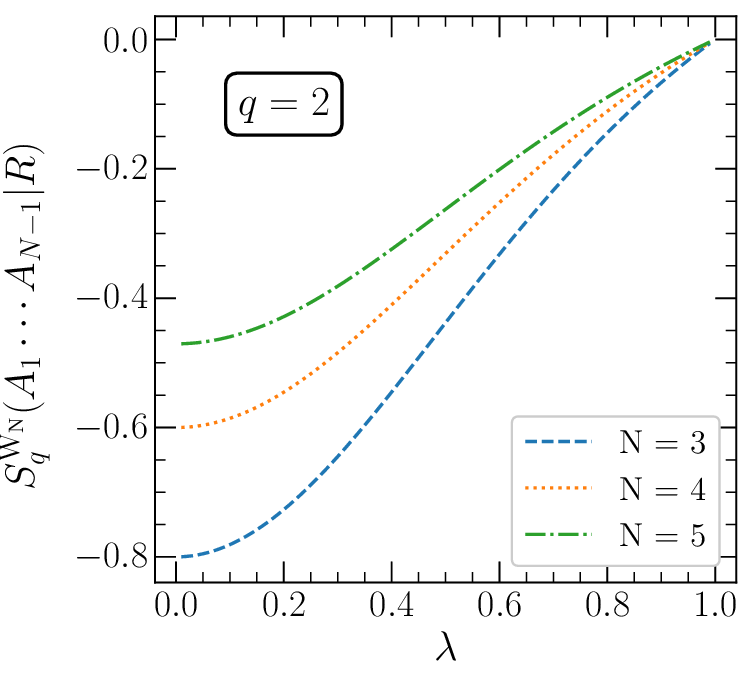}
	\caption{The variation of the AR $ q $-conditional entropy of pure $ N $-qubit W state as a function of the acceleration $ \lambda $ for different values of $ N $ is given in this figure. We have taken $ q=2 $ for this plot.}
	\label{fig:AR_W_n}
\end{figure}

\begin{figure*}[htpb]
	\centering
	\subfloat[]{
		\includegraphics[width=0.32\linewidth]{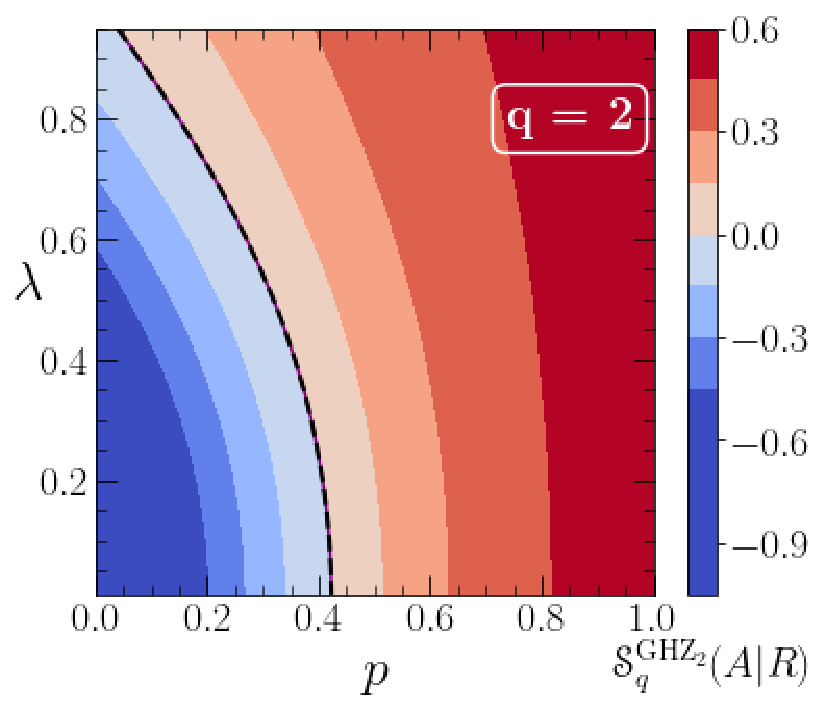}
 		\label{fig:CAR_mixed2qubit_1in1acc_q2}}
	\subfloat[]{
		\includegraphics[width=0.35\linewidth]{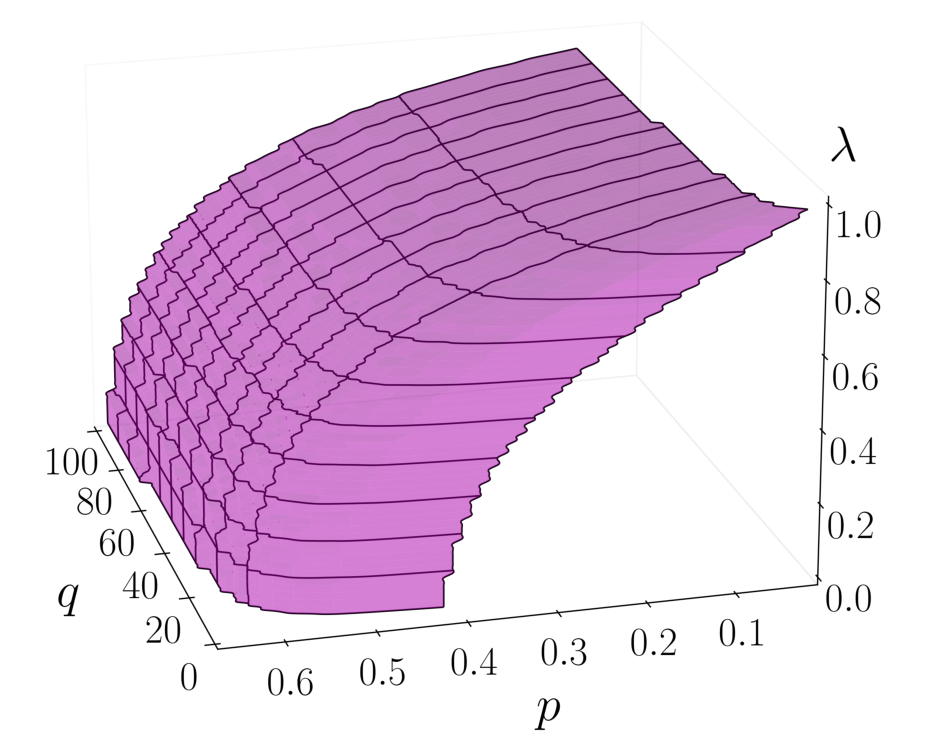}
		\label{fig:CAR_0_mixed2qubit}}
	\subfloat[]{
		\includegraphics[width=0.3\linewidth]{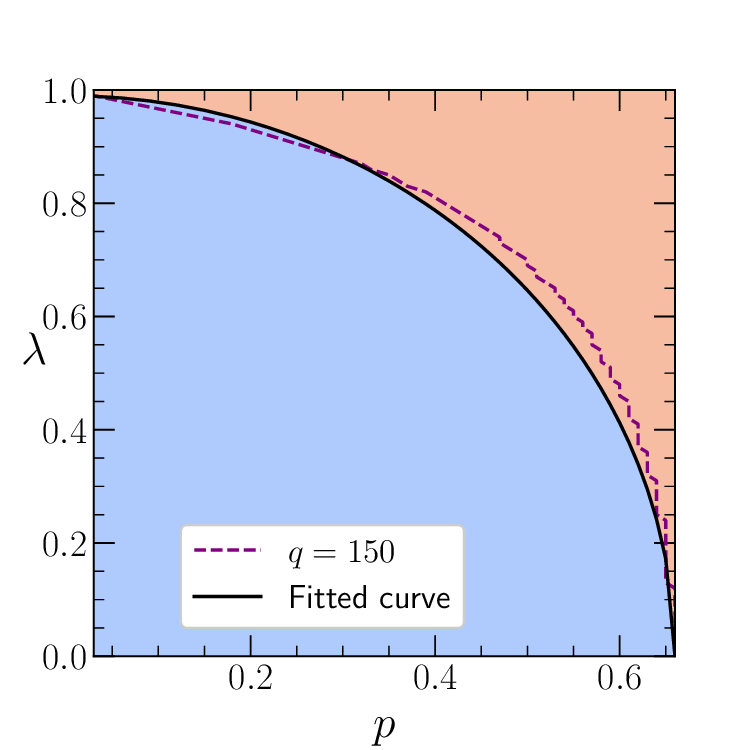}
		\label{fig:CAR_mixed2qubit_lambdavsp}}
	\caption{(a) The variation of the Abe-Rajagopal $ q $-conditional entropy of a single qubit accelerated mixed $ 2 $-qubit GHZ state as a function of $ p $ and $ \lambda $ for $ q = 2 $. The black solid line corresponds to the transition from negative to positive AR $ q $-conditional entropy (b) The variation of the mixing parameter $ p $ for which the $\mathscr{S}^{\text{GHZ}_2}_q(A|R) =0 $, with respect to $ q $ for acceleration parameter $\lambda$. We can observe that the values of the mixing parameter saturate as $ q \rightarrow \infty $ for a given acceleration $\lambda$. (c) The variation of the mixing parameter $ p $ for which the $ \mathscr{S}^{\text{GHZ}_2}_q(A|R) =0 $, with respect to acceleration parameter $ \lambda $ for $ q=150 $. The blue region denotes the region where the state remains non-separable, while the red region denotes the separability.}
\end{figure*}

\subsection{Pure multiqubit W states}
\noindent\textbf{Generalized pure $ 2 $-qubit W state:}
A generalized pure $2$-qubit state $ \ket{{\text{W}_2}} $ shared between Alice and Bob in the inertial frame is given by
\begin{equation}\label{eq:bell2}
	\ket{{\text{W}_2}} =  \cos{\theta}\ket{01} + \sin{\theta}\ket{10}; \; \cos^2{\theta}+\sin^2{\theta} =1.
\end{equation}
Let Bob's qubit be relativistically accelerated with respect to Alice's qubit, and the resulting composite state is $ \rho^{\text{W}_2}_{A R_I R_{II}} $. Then, the mixed density matrix $\rho^{\text{W}_2}_{A R}$, by following the recipe given in Sec. \ref{sec:RQS}, is given by
\begin{align}\label{eq:2qubitw_AR}
	&\rho^{\text{W}_2}_{AR} =\nonumber \frac{1}{\cosh^{2}{r}} \sum_{n} \tanh^{2n} {r}  \\
	&\nonumber \times \Bigg[ \frac{n+1}{\cosh^{2}{r}} \cos^2{\theta} \ket{0n+1}\bra{0n+1} +   \sin^2{\theta} \ket{1n}\bra{1n} \\
	& + \frac{\sqrt{n+1}}{\cosh{r}} \cos{\theta} \sin{\theta} \,\,\Big( \ket{0n+1}\bra{1n} +  \ket{1n} \bra{0n+1}\Big) \Bigg] .	
\end{align}
The structure of this state $ \rho^{\text{W}_2}_{AR} $ is made up of $ 2\times 2 $ blocks along the diagonal, with the remaining off-diagonal elements being zero. This structure is same as Eq.~(\ref{eq:block_w}), however with the non-zero blocks given by 
\begin{equation}
	\Delta_n(\rho^{\text{W}_2}_{AR}) = \tanh^{2n} {r} \begin{pmatrix}	
		\alpha_{n}^2 \cos^2{\theta} & \alpha_n \cos{\theta} \sin{\theta}  \\
		\alpha_n \cos{\theta}  \sin{\theta}  &   \sin^2{\theta}  \\
	\end{pmatrix}.
\end{equation}
Here, $ \alpha_{n} = \frac{\sqrt{n+1}}{\cosh r} $.
The eigenvalues of this $ n^{\text{th}} $ block of the state $ \rho^{\text{W}_2}_{A R} $ are $0$ and 
\begin{equation}\label{eq:pure_2qubit_AR_EV}
	\Lambda_n^\prime = \frac{ \tanh^{2n} r}{\cosh^2 r} \left( \sin^2\theta + \frac{n+1}{\cosh^2 r} \cos^2 \theta\right),
\end{equation} 
satisfying $ \Tr(\rho^{\text{W}_2}_{A R})=\sum_{n=0}^{\infty} \Lambda_n^\prime = 1 $.

The marginal density matrix  $\rho_{R}^{\text{W}_2}$ is computed by tracing out the other subsystem $A$ and is given by 
\begin{align}
	\rho^{\text{W}_2}_{R} \nonumber = &\frac{1}{\cosh^2 r} \sum_{m=0}^{\infty} \tanh^{2m} r \Big[\sin^2 \theta  \ket{m}\bra{m} \\
	&+ \frac{ m+1 }{\cosh^2 r} \cos^2 \theta   \ket{m+1}\bra{m+1} \Big].
\end{align} 
This is an infinite dimensional diagonal matrix whose $ m^{\text{th}} $  eigenvalue is given by 
\begin{equation}\label{eq:pure_2qubit_R_EV}
	\Lambda_m^\prime = \frac{ \tanh^{2m} r}{\cosh^2 r} \left( \sin^2\theta + \frac{m}{\sinh^2 r} \cos^2 \theta\right),
\end{equation}
where $ \Tr(\rho^{\text{W}_2}_{R})=\sum_{m=0}^{\infty} \Lambda_m^\prime = 1 $. 

To characterize the non-separability  of the state of inertial Alice and non-inertial Rob given in Eq.~(\ref{eq:2qubitw_AR}), we substitute the eigenvalues $ \Lambda_n^\prime $'s of $ \rho^{\text{W}_2}_{AR} $ and $ \Lambda_m^\prime $'s of $ \rho^{\text{W}_2}_R $  in the AR $q$-conditional entropy given in Eq.~(\ref{eq:AR1}) to obtain
\begin{align}\label{eq:CAR_pure_2qubit}
	&S^{\text{W}_2}_q(A|R)  \nonumber =\frac{1}{q-1} \\
	& \times \left[ 1 - {\displaystyle \sum_{n=0}^{\infty} \left( \tanh^{2n} r \left(\sin^2{\theta} + \cos^2{\theta} \frac{n+1}{\cosh^2 r}\right)  \right)^q  \over \displaystyle \sum_{m=0}^{\infty} \left( \tanh^{2m} r \left(\sin^2{\theta} + \cos^2{\theta} \frac{m}{\sinh^2 r}\right)\right)^q} \right].
\end{align}
Using the eigenvalue truncation procedure introduced earlier, we chose a sufficiently large $ k\simeq 10^4 $ and numerically computed the AR $ q $-conditional entropy given in Eq.~(\ref{eq:CAR_pure_2qubit}) and characterize the same with respect to the parameter $ q $, the state parameter $\theta$, and Rob's acceleration $ \lambda$. The $ S^{\text{W}_2}_q(A|R) $ always remains less than zero, implying the state remains non-separable for any values of $\lambda$, $\theta$, and $q$.  We observe that irrespective of the value of $ q $, the non-separability of $\rho^{\text{W}_2}_{A R}$ will be high at lower values of acceleration and decreases with the increase in acceleration.  When $ \theta = \pi/4$, Eq.~(\ref{eq:bell2}) corresponds to a $ 2 $-qubit Bell state in an inertial frame and the corresponding $ S^{\text{W}_2}_q(A|R) $ attains maximum non-separability irrespective of the parameter $ q $. These observations are similar to that of the $ 2 $-qubit GHZ state given in Eq.~(\ref{eq:2qubit_AR_ghz}), except the variation with respect to $ \cos^2\theta $ in Fig.~\ref{fig:CAR_pure2qubit_1in1acc_q2_3D} is shifted in case of $\rho_{A R}^{\text{W}_2}$. The characterization of $ S^{\text{W}_2}_q(A|R) $ for $\rho^{\text{W}_2}_{A R}$ in Eq.~(\ref{eq:2qubitw_AR}) with the initial state being the Bell state, with respect to $\lambda$ for various values of $ q $ is same as given in Fig.~\ref{fig:CAR_pure2qubit_inset}.

\noindent\textbf{Generalized pure $ 3 $-qubit W  state:}
A generalized pure $3$-qubit W state, shared between Alice and Bob, is given by 
\begin{equation}\label{eq:3qubit_w}
	\ket{\text{W}_3} = \sin \theta \cos\phi \ket{001} + \sin\theta \sin \phi \ket{010} + \cos \theta \ket{100}.
\end{equation} 
Let $ x= \sin \theta \cos\phi $, $ y =  \sin\theta \sin \phi $ and $ z = \cos \theta$ giving $ x^2 + y^2 +z^2 =1 $, which normalizes the given state for any $\theta$ and $\phi$. Consider a scenario where the first two qubits belong to Alice, and Bob holds the third qubit. Subsequently, as Bob undergoes uniform acceleration (Rob) relative to Alice in an inertial frame, it gives rise to the composite state $ \rho_{A_1 A_2 R_I R_{II}}^{\text{W}_3} $. The final mixed state after tracing out Rob's mode in region II is expressed as 
\begin{align}\label{eq:dm_pure_3w}
	\rho_{A_1A_2R}^{\text{W}_3} \nonumber =  &  \frac{1}{\cosh^2 r}  \sum_{n=0}^{\infty} \tanh^{2n} r \Big[\alpha_n^2 \,\, x^2 \ket{00n+1}\bra{00n+1} \\
		\nonumber & +xz \alpha_n \left(\ket{00n+1} \bra{10n}+\ket{10n} \bra{00 n+1}\right) \\
		\nonumber & + xy \alpha_n \left( \ket{00n+1} \bra{01n}+\ket{01n} \bra{00n+1} \right) \\
		& \nonumber + z^2 \ket{10n} \bra{10n} + y^2 \ket{01n} \bra{01n} \\
		 & +  yz \left(\ket{10n} \bra{01n} + \ket{01n} \bra{10n}\right) \Big].
\end{align} 	
The density matrix of this state $ \rho_{A_1A_2R}^{\text{W}_3} $ assumes a block structure along the diagonal as given in Eq.~(\ref{eq:block_w}) and the remaining off-diagonal elements are zero. However, here, the diagonal blocks are given by
\begin{equation} 
	\Delta_n(\rho_{A_1 A_2 R}^{\text{W}_3})   = \tanh^{2n} r\begin{pmatrix}
		y^2  & y z  & 0 & x y \alpha_{n}\\
		y z  & z^2  & 0 & x z \alpha_{n}\\
		0 & 0 & 0 & 0\\
		x y \alpha_{n} & x z \alpha_{n} & 0 & x^2 \alpha_{n}^2\\
	\end{pmatrix}.
\end{equation}
The non-zero eigenvalues of this $ n^{\text{th}} $ block of the state $ \rho_{A_1A_2R}^{\text{W}_3} $ is given by
\begin{equation}\label{eq:pure_W_A1A2R_EV}
	\delta_n = \frac{\tanh^{2n} r}{\cosh^2 r} \left(1 + \frac{n+1 -\cosh^2 r}{\cosh^2 r} x^2\right).
\end{equation}
The reduced density matrix  $\rho_{R}^{\text{W}_3}$ corresponding to the modes of Rob in the region I of the Rindler spacetime is an infinite dimensional diagonal matrix, with $\delta_m$ as its diagonal elements. Then, the $ m^{\text{th}} $  eigenvalue of $\rho_{R}^{\text{W}_3}$ is given by 
\begin{equation}\label{eq:pure_W_R_EV}
	\delta_m = \frac{\tanh^{2m} r}{\cosh^2 r} \left(1 + \frac{m -\sinh^2 r}{\sinh^2 r} x^2\right). 
\end{equation}
To characterize the non-separability in the state $ \rho_{A_1A_2R}^{\text{W}_3} $ given in Eq.~(\ref{eq:dm_pure_3w}), we substitute these eigenvalues $ \delta_n $'s  and $ \delta_m $'s  in the expression for the AR $q$-conditional entropy for the bipartition  $ A_1A_2:R $ to obtain 
\begin{align}\label{eq:CAR_pure_w}
	&S^{\text{W}_3}_q(A_1A_2|R)  \nonumber =\frac{1}{q-1} \\
	& \times \left[ 1 - {\displaystyle \sum_{n=0}^{\infty} \left( \tanh^{2n} r\left(1 + \frac{n+1 -\cosh^2 r}{\cosh^2 r} x^2\right)  \right)^q  \over \displaystyle \sum_{m=0}^{\infty} \left( \tanh^{2m} r \left(1 + \frac{n -\sinh^2 r}{\sinh^2 r} x^2\right)\right)^q} \right].
\end{align}
We are considering the non-separability in the inertial : non-inertial bipartition, as simultaneous measurements cannot be done in the other possible partitions. To compute this conditional entropy numerically, we use the eigenvalue truncation procedure and get $ k\simeq 10^4 $. Now, using these $k$ eigenvalues of $\rho_{A_1 A_2 R}^{\text{W}_3}$ and $\rho_{R}^{\text{W}_3}$, we numerically compute the AR $ q $-conditional entropy in Eq.~(\ref{eq:CAR_pure_w}) and characterize the same with respect to its state parameters $\theta , \phi$, acceleration $ \lambda $ and parameter $q$. 

We observe that $ S^{\text{W}_3}_q(A_1A_2|R) $ consistently remains below zero, suggesting the non-separability of the state for the entire range of system parameters, acceleration $\lambda$, and parameter $q$. The maximum non-separability for generalized pure $ 3 $-qubit W state occurs for $\theta$ and $\phi$ such that $ \sin \theta \cos \phi = 1/\sqrt{2} $. Moreover, to understand the behavior of $ S^{\text{W}_3}_q(A_1A_2|R) $  with respect to the other numerous parameters involved, we choose a pure $3$-qubit W state with equal coefficients, i.e., with state parameters  $\theta = \cos^{-1}(1/\sqrt{3})$ and $\phi=\pi/4$. Subsequently, the AR $ q $-conditional entropy for this state is plotted as a function of the acceleration parameter $\lambda$ in Fig.~\ref{fig:CAR_pure_w_AB_C_inset} for various $ q $.

When Rob's acceleration is low, the non-separability of the state $\rho_{A_1A_2R}^{\text{W}_3}$ is most pronounced, as indicated by the minimum of $ S^{\text{W}_3}_q(A_1A_2|R) $. However, with increasing acceleration, the conditional entropy rises and approaches  $ 0 $ as $ \lambda $ tends to $1 $, across all values of $ q $. As $ q $ increases, the $ S^{\text{W}_3}_q(A_1A_2|R) $ becomes more negative for low $\lambda$, but converging to $ 0 $ as $ \lambda$ tends to $ 1$. Although the choice of this $\theta$ and $\phi$ are not unique, the behavior of $ S^{\text{W}_3}_q(A_1A_2|R) $ with respect to $ \lambda $ will follow the same trend. However, these values of conditional entropy are higher than the corresponding conditional entropy for the generalized pure $ 3 $-qubit GHZ state with one of its qubit in acceleration, hence indicating generalized pure $ 3 $-qubit W states having weaker non-separability than generalized pure $ 3 $-qubit GHZ states with one of its qubits being accelerated. Similar to the case of pure multiqubit GHZ states, we only consider the pure $N$-qubit W state instead of the generalized pure $N$-qubit W state.

\noindent\textbf{Pure $ N $-qubit W state:}
A pure $ N $-qubit W state, shared between Alice and Bob, is given by
\begin{equation} \label{eq:w}
	\ket{\text{W}_N}  = \frac{\ket{00 \cdots 01}+ \ket{00 \cdots 10} +\cdots  + \ket{10 \cdots 00}}{\sqrt{N}},
\end{equation} 
where the first $ N-1 $ qubits are with Alice, while the last qubit is in Bob's possession. Here, we again consider only the pure $N$-qubit W state instead of the generalized $N$-qubit W state for logistical reasons. The final Alice-Rob state, after accelerating Bob's qubit and tracing out the second Rindler region of Rob's spacetime using the recipe given in Sec. \ref{sec:RQS}, is given by 
\begin{align}\label{eq:Nqubit_W}
	&\rho_{A_1 A_2  \cdots  A_{N-1} R}^{\tiny{\text{W}_N}} = \nonumber \frac{1}{N \cosh^2 r} \sum_{n=0}^{\infty} \tanh^{2n} r \\
	& \nonumber \times \Bigg[ \alpha_n^2 \,\, \ket{00\cdots0n+1}\bra{00 \cdots 0n+1} \\ 
	& \nonumber  +  \Big(\ket{00\cdots1n}+\cdots+\ket{01\cdots0n}+\ket{10\cdots0n}\Big)\\
	& \nonumber \,\,\,\, \times \Big(\bra{00\cdots1n}+\cdots+\bra{01\cdots0n}+\bra{10\cdots0n}\Big) \\
	& \nonumber  + \alpha_n \Big(  \ket{00\cdots0n+1}\big(\bra{00\cdots1n}+\cdots+\bra{01\cdots0n} \\  & +\bra{10\cdots0n}\big) + \text{c.c.} \Big) \Bigg].
\end{align}  
Since one of the qubits is accelerating in a non-inertial frame, a simultaneous measurement cannot be done on the inertial and the non-inertial qubits together.  Hence we proceed our investigations by choosing the $A_1 A_2 \cdots A_{N-1} : R $ bipartition to calculate the AR $ q $-conditional entropy $  S^{\tiny{\text{W}_N}}_q(A_1 A_2 \cdots A_{N-1} | R) $. This conditional entropy is evaluated explicitly for the state given in Eq.~(\ref{eq:Nqubit_W}), using the eigenvalues of the reduced system $\rho_{A_1 A_2 \cdots A_{N-1} R}^{\text{W}_N}$ and its subsystem $\rho_{R}^{\text{W}_N}$ and is given by
\begin{align}\label{eq:nAR_W}
	& S^{\text{W}_N}_q(A_1 A_2 \cdots A_{n-1}|R)  \nonumber = \frac{1}{q-1} \\
	& \times \left[ 1 - { \displaystyle \sum_{n=0}^{\infty} \left(\tanh^{2n} r \left(N-1 +  \frac{n+1}{\cosh^{2} r} \right)\right)^q \over \displaystyle \sum_{m=0}^{\infty} \left(\tanh^{2m} r \left(N-1 +  \frac{m}{\sinh^{2} r} \right)\right)^q}\right].
\end{align}
This conditional entropy $ S^{\text{W}_N}_q(A_1 A_2 \cdots A_{n-1}|R) $ can be numerically computed for various values of $ N $, $ \lambda $, and $ q $ using the eigenvalue truncation procedure given previously. With $k\simeq 10^4$, the $ S^{\text{W}_N}_q(A_1 A_2 \cdots A_{n-1}|R) $ is found to be again non-separable throughout the ranges of system parameters, acceleration $\lambda$ and $q$. The conditional entropy is plotted for various values of $ N $ and $ \lambda $ for $ q=2 $ in Fig.~\ref{fig:AR_W_n}. We observe that irrespective of $ N $, the non-separability of $\rho_{A_1 A_2 \cdots A_{N-1} R}^{\tiny{\text{W}_N}}$ is high at lower values of acceleration and decreases with the increase in acceleration as $ S^{\text{W}_N}_q(A_1 A_2 \cdots A_{n-1}|R) $ tends to $ 0 $ when $ \lambda$ reaches $ 1 $. However, as $ N $ increases, the variation of $ S^{\text{W}_N}_q(A_1 A_2 \cdots A_{n-1}|R) $ with respect to $ \lambda $ becomes less steeper. Hence, we can conclude that the non-separability of the state $\rho_{A_1 A_2 \cdots A_{N-1} R}^{\tiny{\text{W}_N}}$  reduces as the number of qubits inertial subsystem increases. This observation remains true for all values of parameter $ q $.

To consolidate, all the pure multiqubit GHZ and W states with one of its qubits under acceleration will remain non-separable in the inertial : non-inertial bipartition throughout the ranges of its state parameters, parameter $ q $ and acceleration $\lambda$. The states are highly non-separable at low acceleration $\lambda$ and tend towards separability at very high acceleration of $\lambda$ tending to $ 1 $. Moreover, with one of its qubits under acceleration, the non-separability of the pure multiqubit GHZ and W states becomes steeper while approaching $0$ at large acceleration values $\lambda$ and parameter $q$. The non-separability of the pure $N$-qubit GHZ state with one of its qubits being accelerated is independent of the number of qubits, $N$, while for pure $N$-qubit W state, the non-separability decreases with an increase in the number of qubits. Consequently, the pure $N$-qubit W state with one of its qubits under acceleration has weaker non-separability than the pure $N$-qubit GHZ state with one of its qubits under acceleration.

\begin{figure*}[htpb]
	\centering
	\subfloat[]{
		\includegraphics[scale=0.5]{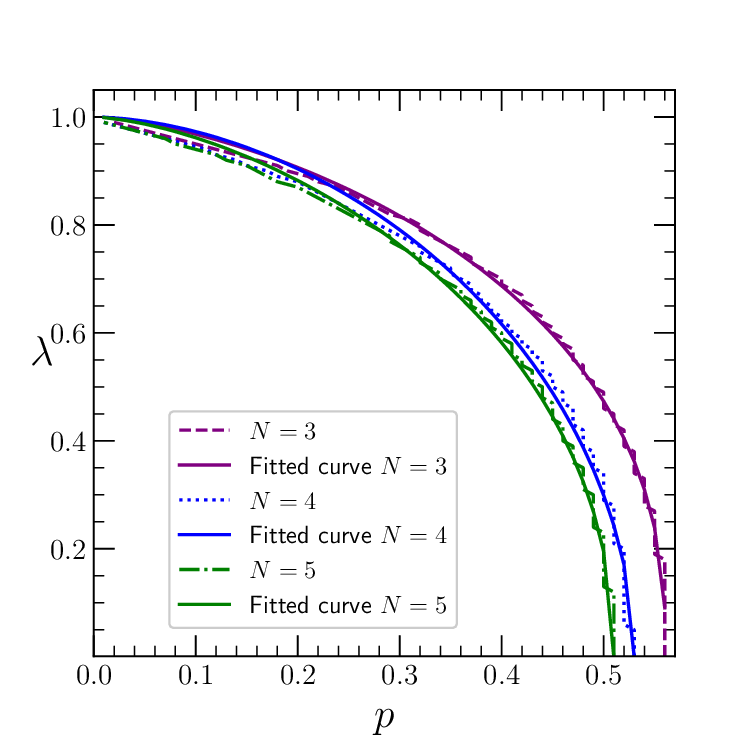}
		\label{fig:CAR_mixedNqubit_ghz_lambdavsp}}
	\subfloat[]{
		\includegraphics[scale=0.5]{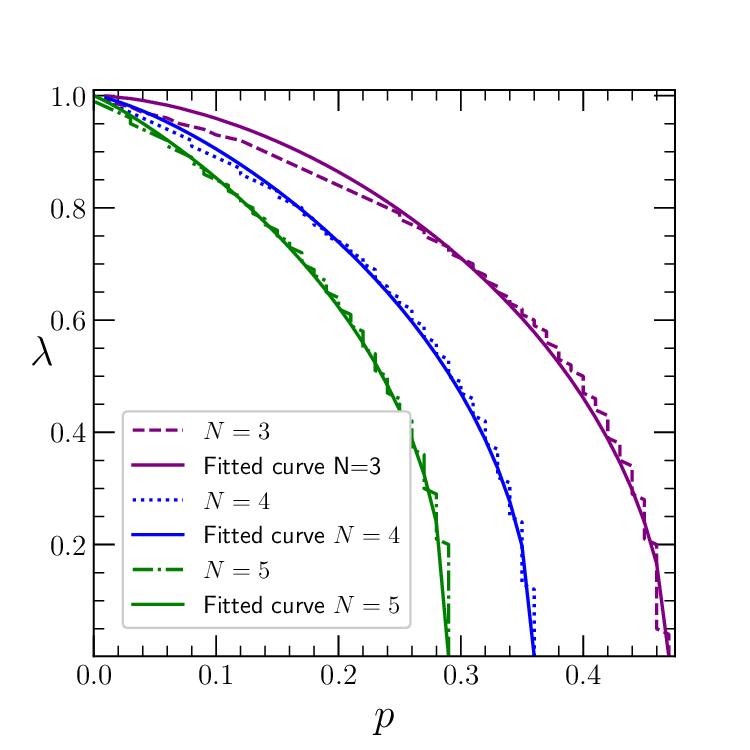}
		\label{fig:CAR_mixedNqubit_w_lambdavsp}}
	\caption{The variation of the non-separable to separable transition points of the AR $ q $-conditional entropy for the (a) mixed $ N $-qubit GHZ state and (b) mixed $ N $-qubit W state, with one of its qubits in uniform acceleration, as a function of the acceleration parameter $ \lambda $ and mixing parameter $ p $ for different values of $ N =3,4 $ and $ 5 $ is given in this figure. The fitted curves are shown in the solid-colored lines. }
\end{figure*}

\section{MIXED MULTIQUBIT STATES WITH AN ACCELERATING QUBIT}\label{sec:mixed}

In this section, we introduce a global noise to all the pure states considered in Sec. \ref{sec:pure} when they have equal probability in their superposed states and study the non-separability of resultant mixed states when one of its qubits is accelerated, using the AR $ q $-conditional entropy. The non-separability of the mixed states with an accelerating qubit thus obtained is characterized with respect to parameter $ q $, acceleration parameter $\lambda$, and mixing parameter $ p $.

\subsection{Mixed multiqubit GHZ states}\label{subsec:dmt}
\noindent\textbf{Mixed  $ 2 $-qubit GHZ state:}
A  mixed $2$-qubit GHZ state shared between Alice and Bob, obtained by adding a global noise to the pure $2$-qubit GHZ state, is given by 
\begin{equation}\label{eq:2qubit_mixedAB}
	\varrho^{\text{GHZ}_2}_{AB} = \frac{p}{4} \mathbb{I}_d + (1-p) \ket{\psi_{AB}}\bra{\psi_{AB}},
\end{equation}
where $ \ket{\psi_{AB}}=  (\ket{00} + \ket{11})/\sqrt{2} $, $ \mathbb{I}_d $ is the $ 4 \times 4 $ identity matrix and $ p $ is the global noise or the  mixing parameter. Note that $ p $ is a parameter in the range $ [0,1] $ and $ p=0 $ denotes the pure $2$-qubit GHZ state, and $ p=1 $ denotes the maximally mixed $ 2 $-qubit GHZ state. Let us consider that Bob is now relativistically accelerating with respect to Alice, and the resulting composite state is $ \varrho^{\text{GHZ}_2}_{AR_IR_{II}} $. We now obtain the reduced state of Alice and Rob in region I by tracing out Rob's mode in region II using the recipe given in Sec. \ref{sec:RQS}, as 
\begin{align}\label{eq:2qubitghz_mixedAR}
	 \nonumber \varrho^{\text{GHZ}_2}_{AR} =& \frac{1}{\cosh^2 r} \sum_{n=0}^{\infty} \tanh^{2n} r \\
	& \nonumber \times \Bigg[\frac{2-p}{4} \left(\ket{0n}\bra{0n} + \frac{n+1}{\cosh^2 r}\ket{1n+1}\bra{1n+1} \right) \\
	&\nonumber + \frac{p}{4}\left(\ket{1n}\bra{1n} + \frac{n+1}{\cosh^2 r}\ket{0n+1}\bra{0n+1}\right) \\
	&+ \frac{1-p}{2} \frac{\sqrt{n+1}}{\cosh r} \Big(\ket{0n}\bra{1n+1} + \ket{1n+1}\bra{0n}\Big)\Bigg].
\end{align}
The Rob's subsystem in region I, by tracing out Alice's qubit, is given by
\begin{align}\label{eq:2qubitghz_mixedR}
	\varrho^{\text{GHZ}_2}_{R} \nonumber = & \frac{1}{2\cosh^2 r} \sum_{n=0}^{\infty} \tanh^{2n} r \\
	& \times \left(\ket{n}\bra{n} + \frac{n+1}{\cosh^2 r}\ket{n+1}\bra{n+1} \right).
\end{align} 
These infinite-dimensional density matrices do not possess a diagonal block structure; hence, arriving at the analytical form for their eigenvalues is not straightforward, and one has to resort to numerical techniques. Below, we describe a procedure for obtaining the eigenvalues of these two density matrices to evaluate the AR $ q $-conditional entropy.

\noindent {\it Density matrix truncation procedure:} To numerically compute the AR $ q $-conditional entropy, we make use of the fact that the sum of the eigenvalues of $\varrho^{\text{GHZ}_2}_{A R}$ and $\varrho^{\text{GHZ}_2}_{R}$ tends to $ 1$ as $ n $ and $ m $ approaches infinity respectively. However, unlike the case of pure states we do not have the analytical expression for $\mathscr{S}_q^{\text{W}_2}(A|R)$ in terms of the eigenvalues of $\varrho^{\text{GHZ}_2}_{A R}$ and $\varrho^{\text{GHZ}_2}_{R}$.  Therefore, we use the numerical technique in which the matrix dimension of infinity is approximated to an agreeable finite dimension $ l \times l $ of the density matrix, such that $ \Tr (\varrho^{\text{GHZ}_2}_{AR}) $ approaches  1. For the density matrix given in Eq.~(\ref{eq:2qubitghz_mixedAR}), the dimension of $\varrho^{\text{GHZ}_2}_{AR}$ for which $ \Tr (\varrho^{\text{GHZ}_2}_{AR}) $ approaches $ 1 $ is found to be around $ l\approx10^3 $. Similarly, eigenvalues are also obtained for Rob's subsystem in the region I, $ \varrho^{\text{GHZ}_2}_{R} $, by limiting its dimension to the finite $l \times l$, such that  $ \Tr (\varrho^{\text{GHZ}_2}_{R}) $ approaches  1.

Now, the AR $ q $-conditional entropy is numerically computed for the state given in Eq.~(\ref{eq:2qubitghz_mixedAR}) using density matrix truncation procedure described above, for various values of Rob's acceleration $ \lambda $, mixing parameter $ p $, and parameter $ q $. The AR $q$-conditional entropy for the initial mixed state whose subsystem is accelerated is given by $ \mathscr{S}^{\text{GHZ}_2}_q(A|R) $. It is plotted as a function of the acceleration $\lambda$ and parameter $ p $ in Fig.~\ref{fig:CAR_mixed2qubit_1in1acc_q2} for $ q=2 $. We can observe that in the plot of $ \mathscr{S}^{\text{GHZ}_2}_q(A|R) $ for the state given in Eq.~(\ref{eq:2qubitghz_mixedAR}), there exist regions of both separability (red shades) and non-separability (blue shades) depending on the choices of $\lambda$ and $p$. The $ \mathscr{S}^{\text{GHZ}_2}_q(A|R) =0$ curve (black dotted) indicates the transition of $\varrho^{\text{GHZ}_2}_{AR}$ from non-separability to separability and it depends on values of $\lambda$ and $p$. The whole transition curve shifts towards the right as $q$ increases, indicating that with an increase in the $q$ value, the area of the non-separability region with respect to $\lambda$ and $p$ keeps increasing. As $q$ is increased to large values, say $50$ and above, the transition curve becomes saturated with respect to $\lambda$ and $p$ as interpreted from Fig.~\ref{fig:CAR_0_mixed2qubit}. Hence, the strongest condition on separability can be obtained in the asymptotic limit of $q$ tending to $\infty$ ($q\rightarrow\infty$). 

For the asymptotic limit of $q \rightarrow \infty$, in Fig.~\ref{fig:CAR_mixed2qubit_lambdavsp}, we plot the transition point from non-separability to separability obtained numerically, i.e., $ \mathscr{S}^{\text{GHZ}_2}_q(A|R) =0$ with respect to $ p $ and $\lambda$ (dashed curve). Note that the saturation occurs beyond $ q=50 $, and we chose a reasonably high value of $ q=150 $ to study separability. The region with blue shade denotes the non-separable states, while the region with red shade denotes the separable states. These transition points for different acceleration can be fitted to a curve of the form
\begin{equation}\label{eq:fit_curve}
	(p/b)^x + \lambda^y =1.
\end{equation}
The choice of the function for the fitted curve is made using the fact that the separability occurs at $ p = b $ when the acceleration $ \lambda =0 $ and $ p = 0  $ as the acceleration $ \lambda \rightarrow 1 $.
Hence, the separability criteria for the state in Eq.~(\ref{eq:2qubitghz_mixedAR}), which relates the mixing parameter $ p $ and Rob's acceleration $ \lambda $, is given by 
\begin{equation}\label{eq:Sep_2qubit}
	p >  \frac{2}{3} \left(1-\lambda^{2.2}\right)^{(1/2.05)}.
\end{equation}
When the Rob's acceleration $\lambda=0$, the necessary and sufficient condition for separability of this state occurs at  $ p>2/3 $ \cite{PhysRevLett.76.722}, which is recovered from the Eq.~(\ref{eq:Sep_2qubit}). As the acceleration increases, the non-separable region keeps reducing, and the non-separability vanishes only at acceleration $\lambda$  approaching $ 1 $. Hence, the Eq.~(\ref{eq:Sep_2qubit}) gives the strongest condition on separability in the asymptotic limit of $q$ for the mixed $ 2 $-qubit GHZ state with one of its qubits being accelerated.

\begin{table*}
	\caption{This table shows the conditions for separability obtained using the AR $ q $-conditional entropy for multiqubit GHZ and W states. The pure multiqubit states with one of its qubits in acceleration remain non-separable (NS) everywhere. The conditions for mixed GHZ and W states fit the family of curves given in Eq.~(\ref{eq:fit_curve}).}
	\label{Tab:separability}
	\begin{ruledtabular}\centering
		\begin{tabular}{c|c|c|c}
			N & Pure $\rho^{\text{GHZ}_N}_{AR}/\rho^{\text{W}_N}_{AR}$ &  mixed $\varrho_{AR}^{\text{GHZ}_N}$ & mixed $\varrho_{AR}^{\text{W}_N}$\\ \hline 
			2 & NS &$  \frac{2}{3} \left(1-\lambda^{2.2}\right)^{(1/2.05)} $ & $  \frac{2}{3} \left(1-\lambda^{1.57}\right)^{(1/2.6)} $ \\
			3 & NS & $ p > 0.56 (1 -  \displaystyle \lambda^{2.47})^{ (1/1.49)} $ & $p >  0.47  (1 -\lambda^{1.86})^{ (1/1.68)} $ \\
			4 &NS& $ p > 0.53 (1 -  \lambda^{1.92})^{ (1/1.78)}$ & $p >  0.36  (1 -\lambda^{2.04})^{ (1/1.32)} $ \\
			5 & NS& $ p > 0.51 (1 -\lambda^{2.14})^{ (1/1.55)}$ & $p >  0.29  (1 -\lambda^{2.28})^{ (1/1.12)} $ 
		\end{tabular}
	\end{ruledtabular}
\end{table*}

\noindent\textbf{Mixed $ N $-qubit GHZ state:}
We add a global noise to pure $ N $-qubit GHZ state to obtain a mixed $N$-qubit GHZ state, which is shared between two parties, Alice and Bob, in the inertial frame, where Alice has the first $ N-1 $ qubits, and Bob has $ N^{\text{th}} $ qubit in his possession. This state is given by
\begin{align}\label{eq:Nqubit_mixedAB_ghz}
	\varrho^{\text{GHZ}_N}_{A_1 A_2 \cdots A_{N-1}B}  &\nonumber  = \frac{p}{2^N} \mathbb{I}_d \\
	&+ (1-p) \ket{\psi_{A_1 A_2 \cdots A_{N-1}B}}\bra{\psi_{A_1 A_2 \cdots A_{N-1}B}},
\end{align}
where $ \ket{\psi_{A_1 A_2 \cdots A_{N-1}B}}= (\ket{0}^{\otimes N}+ \ket{1}^{\otimes N})/\sqrt{2}  $, $ \mathbb{I}_d $ is the $ 2^N \times 2^N $ identity matrix and $ p $ is the mixing parameter. Let us now consider that Bob is uniformly accelerating with respect to Alice and the resulting state $ \varrho_{A_1 A_2 \cdots A_{N-1}R_IR_{II}}^{\text{GHZ}_N} $ is obtained by using the transformations from Minkowski to Rindler spacetime in Eqs.~(\ref{eq:R0}) and (\ref{eq:R1}). The final state $ \varrho_{A_1 A_2 \cdots A_{N-1}R}^{\text{GHZ}_N}  $ shared between Alice and Rob's mode in the region I is obtained using the recipe given in Sec. \ref{sec:RQS} and is given by
\begin{widetext}
\begin{align}\label{eq:Nqubit_mixedAR}
	\varrho_{A_1 A_2 \cdots A_{N-1}R}^{\text{GHZ}_N} & \nonumber= \frac{1}{\cosh^2 r} \sum_{n=0}^{\infty}\tanh^{2n} r \\
	& \nonumber\Bigg[  \frac{2-p}{2^N} \Big(  \left(\ket{00\cdots n}+\ket{01\cdots n} + \cdots+\ket{11\cdots n}\right) \left(\bra{00\cdots n}+\bra{01\cdots n}+\cdots+\bra{11\cdots n}\right)\\
	& \nonumber + \alpha_{n}^2 \left(\ket{00\cdots n+1}+\ket{01\cdots n+1}+\cdots+\ket{11\cdots n+1}\right) \left(\bra{00\cdots n+1}+\bra{01\cdots n+1}+\cdots+\bra{11\cdots n+1}\right) \Big) \\
	& \nonumber + \frac{1-p}{2} \Bigg( \ket{00 \cdots 0n} \bra{00 \cdots 0n}  +  \alpha_{n} \Big(  \ket{00 \cdots 0n} \bra{11 \cdots 1 n+1} +  \ket{11 \cdots 1 n+1}\bra{00 \cdots 0n} \Big)\\
	&  + \alpha_{n}^2 \ket{11 \cdots 1 n+1} \bra{11 \cdots 1 n+1}\Bigg) \Bigg].
\end{align}
\end{widetext}
The Rob's subsystem is obtained by tracing out the inertial Alice's qubits, i.e.,
\begin{equation*}
	\varrho_{R}^{\text{GHZ}_N} = \Tr_{A_1 A_2 \cdots A_{N-1}} (\varrho_{A_1 A_2 \cdots A_{N-1}R}^{\text{GHZ}_N}).
\end{equation*} 
Using the density matrix truncation procedure given in the last subsection, we can numerically obtain the AR $ q $-conditional entropy $ \mathscr{S}_q^{\text{GHZ}_N}(A_1 A_2  \cdots  A_{N-1}|R) $ in the inertial : non-inertial bipartition for any choice of $N$. For example, when a mixed $ 3 $-qubit GHZ state with one of its qubits under acceleration is considered (see Appendix \ref{app:GHZ3}), we observe a transition from non-separability to separability depending on the values of acceleration $ \lambda $, parameter $ q $ and mixing parameter $ p $. In the asymptotic limit of $ q \rightarrow \infty$, the transition from non-separability to separability saturates to a curve, which is fitted to the function given in Eq.~(\ref{eq:fit_curve}).  Then, the separability criteria for the state $ \varrho_{A_1A_2R}^{\text{GHZ}_3} $ (see Eq.~(\ref{eq:3qubit_mixedA1A2R_ghz})), in terms of the mixing parameter, is given as
\begin{equation}
	p > 0.56 (1 -  \lambda^{2.47})^{ (1/1.49)} .
\end{equation}  
Here, the separability occurs at $ p = 0.56 $ when the acceleration $\lambda=0$ and it reduces to $ p=0  $ when $\lambda=1$ indicating the region of non-separability decreases with increase in acceleration for the $ 3 $-qubit mixed GHZ state with one of its qubits being accelerated. 

A similar investigation is performed for mixed multiqubit GHZ states with $ N=4,5 $ when one of their qubits is accelerated. The non-separability to separability transition with respect to $\lambda$ and $ p $ in the asymptotic limit of $ q\rightarrow\infty $ is observed to be similar to the $ N=3 $ case. To study the impact of $ N $ on the non-separability to separability transition, we plot $ \mathscr{S}^{\text{GHZ}_N}_q( A_1A_2 \cdots A_{N-1}|R) =0$ with respect to $ p $ and $\lambda$  for $ N =3,4$ and $5$ in Fig.~\ref{fig:CAR_mixedNqubit_ghz_lambdavsp}. The strongest condition on separability obtained in the asymptotic limit of $ q \rightarrow\infty$ for these states are fitted to the family of curves in Eq.~(\ref{eq:fit_curve}). The non-separability region of mixed $ N $-qubit GHZ states with one of its qubits under acceleration reduces as $ N $ increases. The states tend to become separable for all choices of $ N $ as the acceleration $\lambda$ approaches  $ 1 $. These separability conditions for the different mixed $N$-qubit GHZ states with one of their qubits being accelerated $ (N=2,3,4$ and $5) $ are consolidated in table \ref{Tab:separability}.

\subsection{Mixed multiqubit W states}
\noindent\textbf{Mixed  $ 2 $-qubit W state:}
A mixed $2$-qubit state shared between Alice and Bob, obtained by introducing a global noise to the pure $ 2 $-qubit W state, is given by 
\begin{equation}\label{eq:2qubit_mixedAB}
	\varrho^{\text{W}_2}_{AB} = \frac{p}{4} \mathbb{I}_d + (1-p) \ket{\psi_{AB}}\bra{\psi_{AB}},
\end{equation}
where $ \ket{\psi_{AB}}=  (\ket{01} + \ket{10})/\sqrt{2} $, $ \mathbb{I}_d $ is the $ 4 \times 4 $ identity matrix and $ p $ is the mixing parameter. Let us consider that Bob is now relativistically accelerating with respect to Alice, and the resulting composite state is $ \varrho_{AR_IR_{II}}^{\text{W}_2} $. We now obtain the reduced state of  Alice and Rob in region I by tracing out Rob's mode in region II using a recipe given in Sec. \ref{sec:RQS}, as 
\begin{align}\label{eq:2qubit_mixedAR_w}
	\nonumber\varrho_{AR}^{\text{W}_2}=&  \frac{1}{\cosh^2 r} \sum_{n=0}^{\infty} \tanh^{2n} r  \\
	 & \nonumber\times \Bigg[ \frac{2-p}{4} \left(\ket{1n}\bra{1n} + \frac{n+1}{\cosh^2 r}\ket{0n+1}\bra{0n+1} \right) \\
	&\nonumber + \frac{p}{4}\left(\ket{0n}\bra{0n} + \frac{n+1}{\cosh^2 r}\ket{1n+1}\bra{1n+1}\right) \\
	&+\frac{1-p}{2} \frac{\sqrt{n+1}}{\cosh r} \Big(\ket{0n+1}\bra{1n} + \ket{1n}\bra{0n+1}\Big)\Bigg].
\end{align}
The Rob's subsystem $ \varrho_{R}^{\text{W}_2} $ is obtained by tracing out the inertial Alice's qubits from the state $ \varrho_{AR}^{\text{W}_2} $ and is same as $ \varrho_{R}^{\text{GHZ}_2} $  given in Eq.~(\ref{eq:2qubitghz_mixedR}). Now, using the density matrix truncation procedure given in Sec. \ref{subsec:dmt}, the conditional entropy $ \mathscr{S}^{\text{W}_2}_q(A|R) $ can be numerically evaluated for this state. The characterization of $\mathscr{S}^{\text{W}_2}_q(A|R) $ for this state qualitatively remains the same as given in Figs.~\ref{fig:CAR_mixed2qubit_1in1acc_q2} and \ref{fig:CAR_0_mixed2qubit} for mixed $ 2 $-qubit GHZ state with one of its qubit in acceleration, but differ in values. The non-separability to separability transition for this state in the asymptotic limit of $ q $ can also be fitted to the function given in Eq.~(\ref{eq:fit_curve}). Thus, the separability criteria in terms of mixing parameter and acceleration $\lambda$  for $\rho^{\text{W}_2}_{AR}$ is given by
\begin{equation}
	 p > \frac{2}{3} \left(1-\lambda^{1.57}\right)^{(1/2.6)}, 
\end{equation}
indicating when $\lambda=0$, we get $ p>2/3 $ as the separable region and the separable region increases as $\lambda$ increases. The mixing parameter $p$ above which the state is separable tends to $0$ as the acceleration $\lambda $ tends to $1$ in the asymptotic limit of $ q $.

\noindent\textbf{Mixed  $ N $-qubit W state:}
We get a mixed $N$-qubit W state by introducing a global noise to the pure $ N $-qubit W state, which is given by
\begin{align}\label{eq:Nqubit_mixedAB_w}
	\varrho^{\text{W}_N}_{A_1 A_2  \cdots  A_{N-1}B}  &\nonumber  = \frac{p}{2^N} \mathbb{I}_d \\
	&+ (1-p) \ket{\psi_{A_1 A_2  \cdots  A_{N-1}B}}\bra{\psi_{A_1 A_2  \cdots  A_{N-1}B}},
\end{align}
where $ \ket{\psi_{A_1 A_2  \cdots  A_{N-1}B}}= (\ket{00 \cdots 01}+ \ket{00 \cdots 10} + \cdots +  \ket{10 \cdots 00})/\sqrt{N} $, $ \mathbb{I}_d $ is the $ 2^N \times 2^N $ identity matrix and $ p $ is the mixing parameter. In the inertial frame, this state is shared between the two parties, Alice and Bob, where Alice has the first $ N-1 $ qubits, and Bob has $ N^{\text{th}} $ qubit in his possession.
Let us now consider that Bob is now uniformly accelerated with respect to Alice and the resulting state $ \varrho_{A_1 A_2  \cdots  A_{N-1}R_IR_{II}}^{\text{W}_N} $ is obtained by using the transformations from Minkowski to Rindler spacetime given in Eqs.~(\ref{eq:R0}) and (\ref{eq:R1}). The final state $ \varrho_{A_1 A_2  \cdots  A_{N-1}R}^{\text{W}_N}  $ between Alice and modes of Rob in the region I using the recipe given in Sec. \ref{sec:RQS} is given by
\begin{widetext}
	\begin{align}\label{eq:Nqubit_w_mixedAR}
		\varrho_{A_1 A_2  \cdots  A_{N-1}R}^{\text{W}_N}   & \nonumber= \frac{1}{\cosh^2 r} \sum_{n=0}^{\infty} \tanh^{2n} r \\
		& \nonumber\Bigg[  \frac{2-p}{2^N} \Big( \left(\ket{00 \cdots n}+\ket{01 \cdots n}+ \cdots +\ket{11 \cdots n}\right) \left(\bra{00 \cdots n}+\bra{01 \cdots n}+ \cdots +\bra{11 \cdots n}\right)\\
		& \nonumber + \alpha_n^2 \left(\ket{00 \cdots n+1}+\ket{01 \cdots n+1}+ \cdots +\ket{11 \cdots n+1}\right) \left(\bra{00 \cdots n+1}+\bra{01 \cdots n+1}+ \cdots +\bra{11 \cdots n+1}\right) \Big) \\
		& \nonumber + \frac{1-p}{N} \Bigg( \alpha_n^2 \,\, \ket{00\cdots 0n+1}\bra{00\cdots 0n+1} +  \alpha_n \Big( \ket{00\cdots 0n+1}\Big(\bra{00\cdots 1n}+ \cdots +\bra{10\cdots0n}\Big) \\ & \nonumber  +  \Big(\ket{00\cdots1n}+ \cdots +\ket{10\cdots0n}\Big) \bra{00\cdots0n+1} \Big) \\ &    +   \Big(\ket{00\cdots1n}+ \cdots +\ket{10\cdots0n}\Big)  \Big(\bra{00\cdots1n}+ \cdots +\bra{10\cdots0n}\Big)\Bigg) \Bigg].
	\end{align}
\end{widetext}
The Rob's subsystem $ \varrho_{R}^{\text{W}_N} $ is obtained by tracing out the inertial Alice's qubits. We can numerically obtain the AR $ q $-conditional entropy $ \mathscr{S}_q^{\text{W}_N}(A_1 A_2  \cdots  A_{N-1}|R) $ in the inertial : non-inertial bipartition, using the density matrix truncation procedure given in Sec. \ref{subsec:dmt} for any choice of $N$. For example, when $ N=3 $, a mixed $ 3 $-qubit W state with one of its qubits under acceleration is considered (see Appendix \ref{app:W3}), depending on the values of acceleration $ \lambda $, parameter $ q $ and mixing parameter $ p $, we observe a transition from non-separability to separability. This transition saturates to a curve in the asymptotic limit $ q \rightarrow \infty$, which is fitted to the function given in Eq.~(\ref{eq:fit_curve}).  Then, the separability criteria for the state $ \varrho_{A_1 A_2R}^{\text{W}_3} $ given in Eq.~(\ref{eq:3qubit_mixedA1A2R_w}), in terms of the mixing parameter $ p $ and $\lambda$, is 
\begin{equation}
	p >  0.47  (1 -\lambda^{1.86})^{ (1/1.68)},
\end{equation}  
implying that $ p>0.47$ is a separable region when  $\lambda=0$ and as $\lambda$ tends to $ 1 $, the state remains separable irrespective of the value of mixing parameter $ p $. 

Further, we observe a similar non-separability to separability transition with respect to $\lambda$ and $ p $ in the asymptotic limit of $ q\rightarrow\infty $ for $ N=4 $ and $ 5 $. In Fig.~\ref{fig:CAR_mixedNqubit_w_lambdavsp}, we plot $ \mathscr{S}^{\text{GHZ}_N}_q( A_1A_2 \cdots A_{N-1}|R) =0$ with respect to $ p $ and $\lambda$  for different values of $ N =3,4$ and $5$. These separability to non-separability transitions are fitted to the family of curves in Eq.~(\ref{eq:fit_curve}). We can observe that the non-separability region reduces as $ N $ increases. Moreover, this reduction is much faster than its mixed $ N $-qubit GHZ counterpart, indicating that mixed $ N $-qubit GHZ states can sustain non-separability better than mixed $ N $-qubit W states when one of their qubits is uniformly accelerated with respect to others. The consolidated conditions of separability of single qubit accelerated mixed multiqubit GHZ and W states are presented in Table \ref{Tab:separability}.

\begin{figure*}[htpb]
	\centering
	\subfloat[]{
		\includegraphics[scale=0.4]{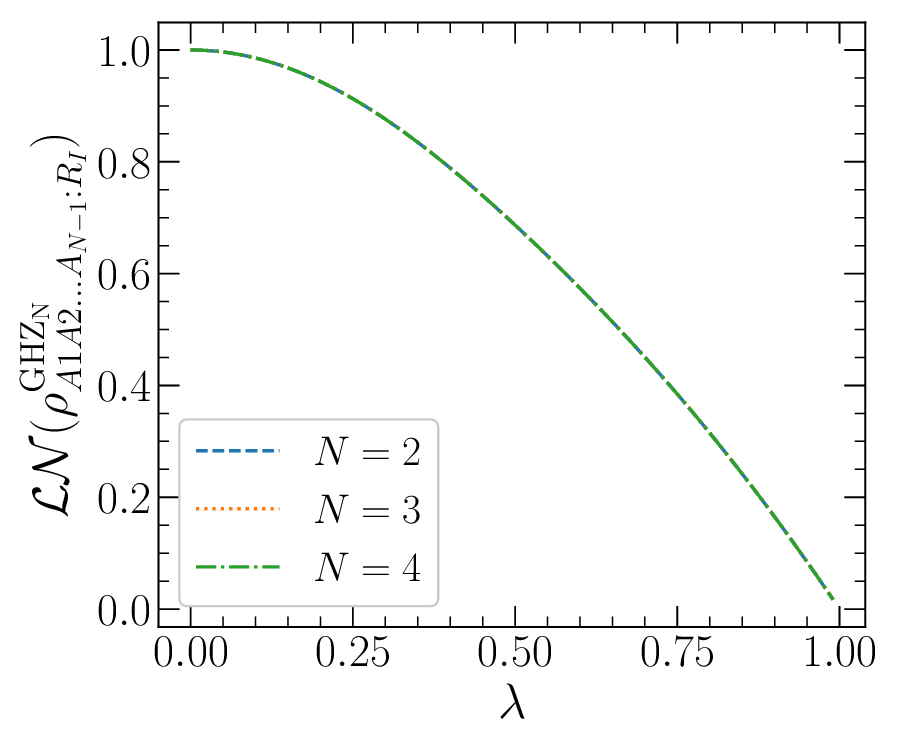}
		\label{fig:neg_pureNqubit_ghz}}
	\subfloat[]{
		\includegraphics[scale=0.4]{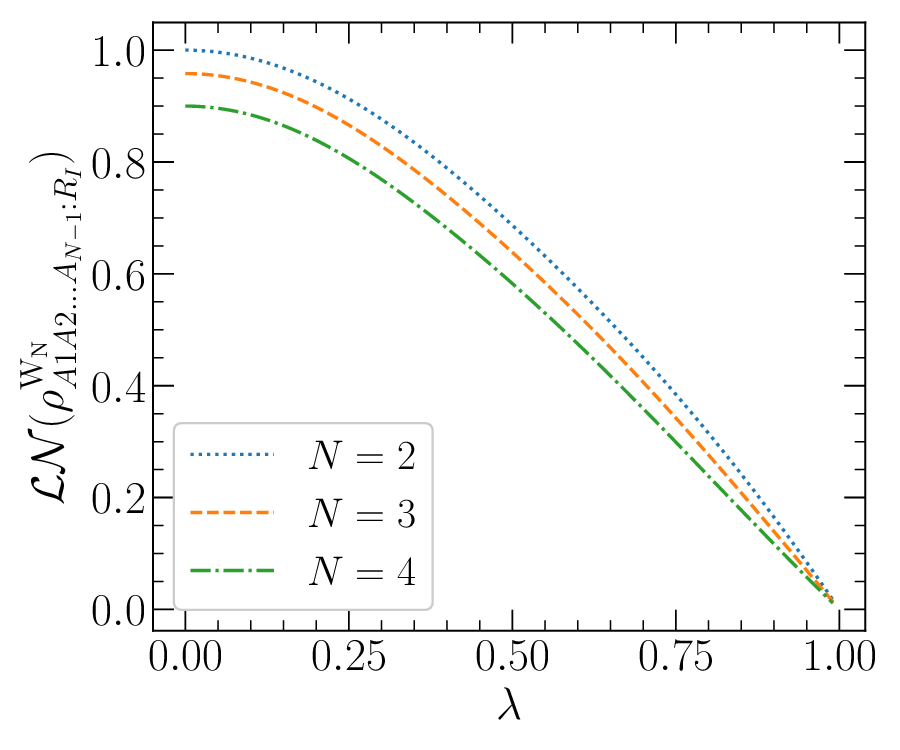}
		\label{fig:neg_pureNqubit_w}}
	\caption{The variation of the logarithmic negativity of single qubit accelerated pure $ N $-qubit (a) GHZ state and (b) W state as a function of the acceleration $ \lambda $ for different number of qubits $ N $ in these states. }
\end{figure*}

\begin{figure*}[htpb]
	\centering
	\subfloat[]{
		\includegraphics[width=0.34\linewidth]{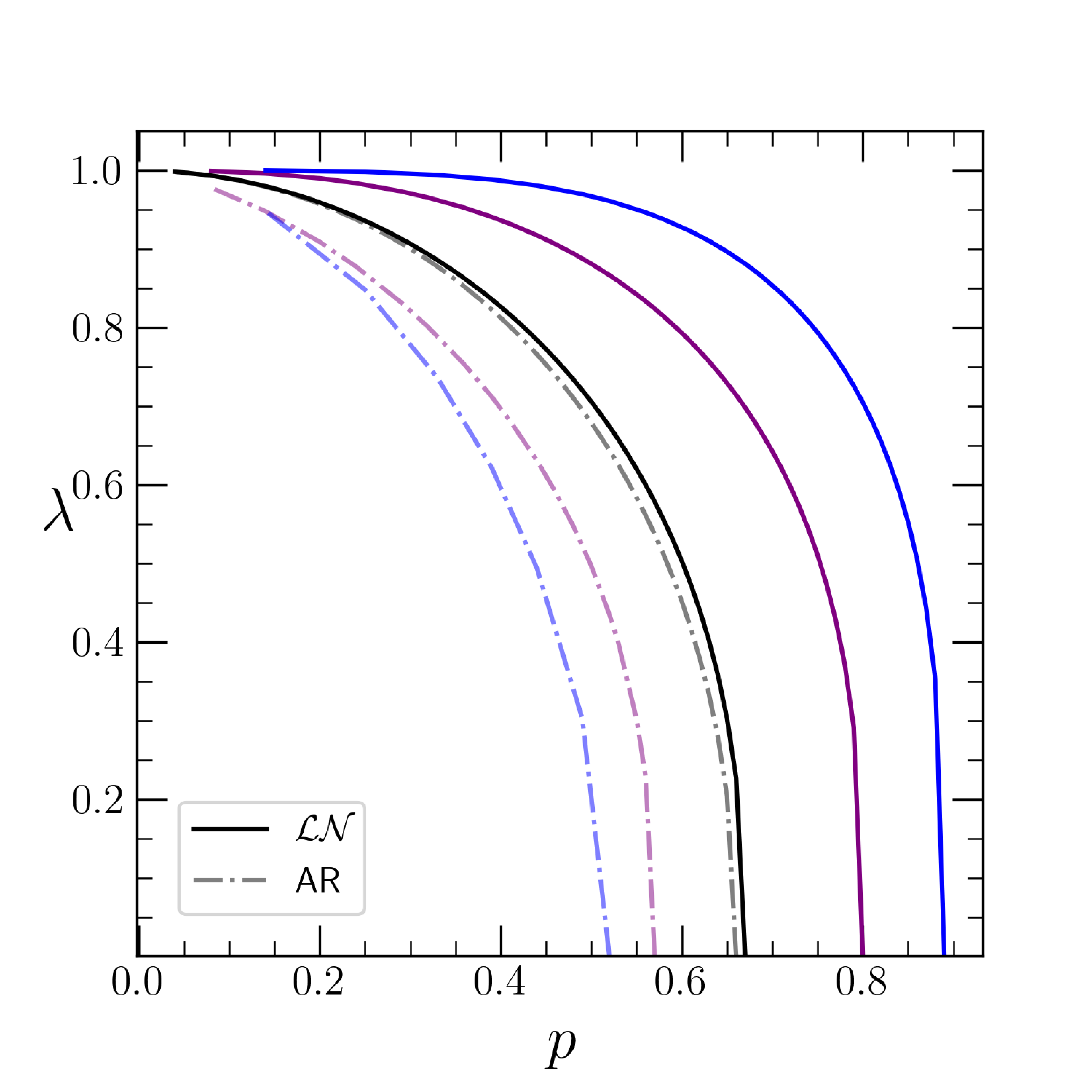}
		\label{fig:neg_mixedNqubit_ghz_lambdavsp}}
	\subfloat[]{
		\includegraphics[width=0.34\linewidth]{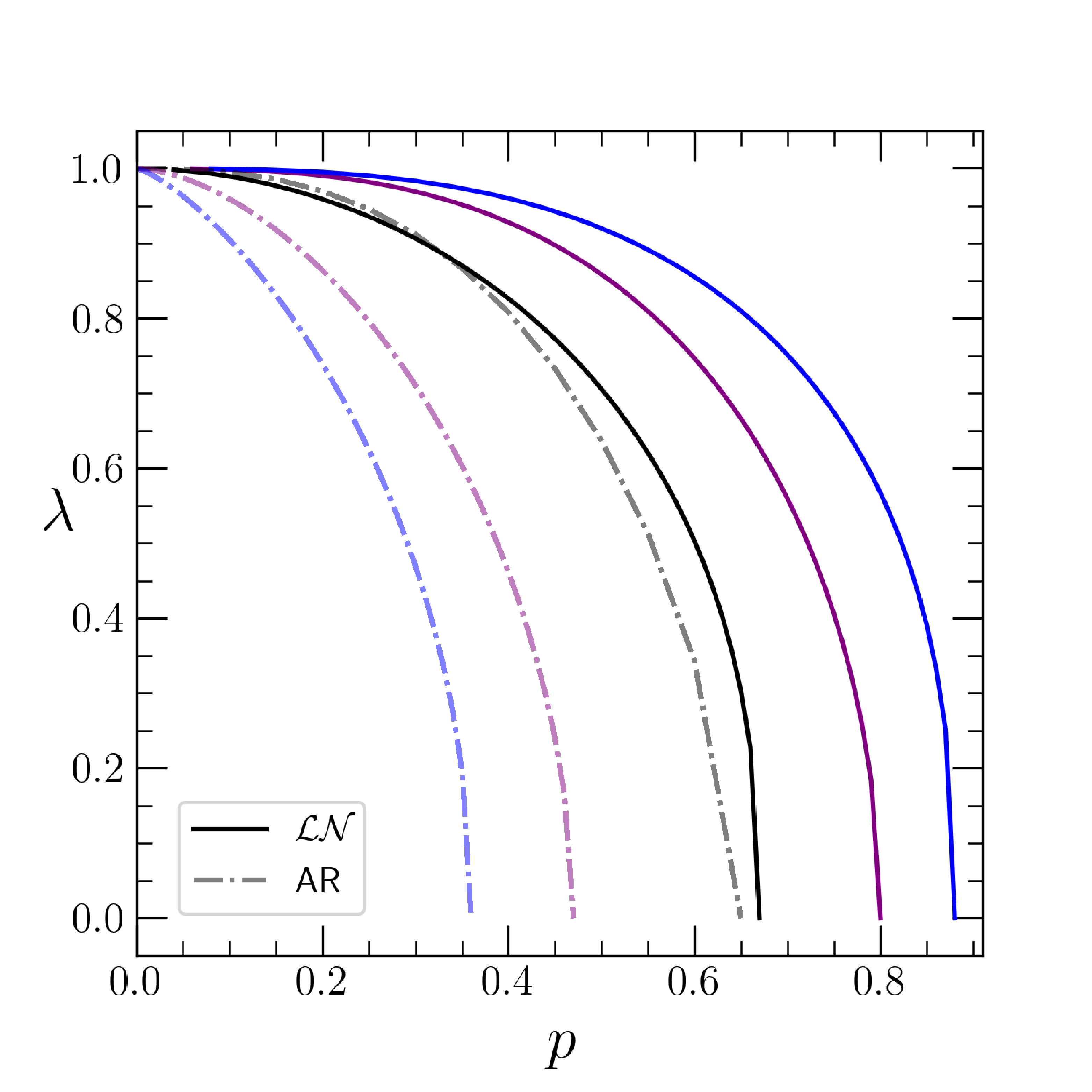}
		\label{fig:neg_mixedNqubit_w_lambdavsp}}
	\caption{Transition from non-separability to separability using AR $q$-conditional entropy (dash-dotted curves) and entanglement to separability using the logarithmic negativity (solid curves) for the single qubit accelerated $ N $-qubit mixed (a) GHZ state and (b)  W state as a function of the acceleration parameter $ \lambda $ and mixing parameter $ p $. Here, the number of qubits in the system are $N=2$ (black), $3$ (violet) and $4$ (blue).    }
\end{figure*}

\section{AR $q$-conditional entropy and Logarithmic Negativity}\label{sec:LN}
We now draw a comparison between the non-separability obtained through AR $q$-conditional entropy and the entanglement measure, viz., logarithmic negativity \cite{plenio2005logarithmic}.  Logarithmic negativity for the state $\rho_{A R_I}$ in the inertial : non-inertial bipartition can be defined as 
\begin{equation}
	{\mathcal{LN}} (\rho_{A:R_{I}}) = \log_2 \norm{\rho^{T_{R_{I}}}_{AR_{I}}}_1,
	\label{eq:LN1}
\end{equation}  
where $T_{R_{I}}$ is the partial transpose taken over the subsystem $\rho_{R_I}$ and $\norm{.}$ gives the trace norm. $\mathcal{LN} (\rho_{AB}) = 0$ for separable states and $\mathcal{LN} (\rho_{AB}) > 0$ for entangled states. Further, the definition of logarithmic negativity can be extended to single qubit accelerated $N$-qubit state as
\begin{equation}
	{\mathcal{LN}} (\rho_{A_1A_2 \cdots A_{N-1}:R_{I}}) = \log_2 \norm{\rho^{T_{R_{I}}}_{A_1A_2 \cdots A_{N-1}R_{I}}}_1,
	\label{eq:LN_n}
\end{equation}
where the bipartition is considered between the inertial Alice's particles, $A_1,A_2, \cdots, A_{N-1}$, and non-inertial Rob's mode $R_I$. 

It has been shown that the logarithmic negativity for a single qubit accelerated maximally entangled $2$-qubit state degrades as acceleration increases and it tends to zero in the infinite acceleration limit \cite{PhysRevLett.95.120404}. Here, we characterize the behavior of logarithmic negativity (see Eq.~(\ref{eq:LN_n})) in a single qubit accelerated $N$-qubit pure GHZ and W states given in Eqs.~(\ref{eq:NGHZ_AR}) and (\ref{eq:Nqubit_W}) respectively. The variation of logarithmic negativity for a single qubit accelerated $N$-qubit pure GHZ and W states (for $N=2,3,$ and $4$) with respect to the acceleration $\lambda$ is plotted in Figs.~\ref{fig:neg_pureNqubit_ghz} and \ref{fig:neg_pureNqubit_w} respectively. The entanglement for a single qubit accelerated $N$-qubit pure GHZ state is independent of the number of qubits in the system, whereas entanglement decreases as the number of qubits increases for a single qubit accelerated $N$-qubit pure W states. However, the entanglement in both these pure multiqubit states reduces as acceleration $\lambda$ increases and goes to zero at infinite acceleration.  The AR $q$-conditional entropy of single particle accelerated $N$-qubit pure GHZ and W states also show the same behavior for non-separability, such that the non-separability reduces as acceleration increases, and the state tends towards separability in the infinite acceleration limit.

The logarithmic negativity for a single qubit accelerated $2$-, $3$-, and $N$-qubit mixed GHZ state (given in Eqs.~(\ref{eq:2qubitghz_mixedAR}), (\ref{eq:3qubit_mixedA1A2R_ghz}), and (\ref{eq:Nqubit_mixedAR})) and W state (given in Eqs.~(\ref{eq:2qubit_mixedAR_w}), (\ref{eq:3qubit_mixedA1A2R_w}) and (\ref{eq:Nqubit_w_mixedAR}) in the inertial: non-inertial bipartition can be evaluated using its definition given in  Eq.~(\ref{eq:LN_n}). The entanglement to separability transition in these states as a function of the acceleration and mixing parameter $p$ is given in Fig.~\ref{fig:neg_mixedNqubit_ghz_lambdavsp} and \ref{fig:neg_mixedNqubit_w_lambdavsp}. 
For any $N$, all multiqubit mixed states for which the logarithmic negativity lies under the transition curve represent the entangled states, whereas those with logarithmic negativity above the transition curve represent the separable states. For $N=2$, the transition curves obtained from AR $q$-conditional entropy and logarithmic negativity are equivalent for both mixed GHZ and W states. However, as $N$ increases, the entanglement to separability transition from logarithmic negativity is obtained at higher mixing parameter $p$ for a particular acceleration $\lambda$,
whereas the non-separability to separability transition from AR $ q $-conditional entropy occurs at a lower $ p $ for the same $\lambda$ (see Figs.~\ref{fig:CAR_mixedNqubit_ghz_lambdavsp} and \ref{fig:CAR_mixedNqubit_w_lambdavsp}). Strikingly, the entanglement (non-separability) provided by the logarithmic negativity (AR $q$-conditional entropy) for the single qubit accelerated $N$-qubit mixed GHZ state is stronger than that of the single qubit accelerated $N$-qubit mixed W state. Therefore, we can conclude that the AR $q$-conditional entropy of single particle accelerated mixed multiqubit GHZ and W states provide a more robust condition on non-separability than the logarithmic negativity as the number of qubits in the system increases.

\section{Conclusions}\label{sec:conclusion}

In an inertial frame, a shared state between two parties, Alice and Bob, $\rho_{A B}$, can be represented using Minkowski coordinates. When Bob undergoes a uniform acceleration, the vacuum and excited states in  Minkowski coordinates  transform into the corresponding two-mode squeezed vacuum states in Rindler coordinates. The two modes of this accelerated Bob's (Rob) state belong to the two disjoint regions of the Rindler spacetime. Therefore, the final mixed state of inertial Alice and non-inertial Rob, $\rho_{A R}$, can be obtained by tracing out the region II of the Rindler spacetime. The non-separability of this relativistic reduced state is characterized using the AR $ q $-conditional entropy, which takes negative values for non-separable states and positive values for separable states. We captured the non-separability of pure multiqubit GHZ and W states when one of their qubits is accelerated. Further, we studied the non-separability of mixed states generated by adding a global noise to the same multiqubit GHZ and W states when one of their qubits is in uniform acceleration.

The analytical form of the AR $q$-conditional entropy for pure multiqubit GHZ and W states is evaluated using eigenvalues of the state of inertial Alice and Rob's mode in region I ($\rho_{A_1A_2 \cdots A_{N-1} R}$) and the Rob's subsystem ($\rho_{R}$). We then numerically studied the non-separability using the {\it eigenvalue truncation procedure} (see Sec. \ref{subsec:evt}) performed on these infinite dimensional density matrices  $\rho_{A_1A_2 \cdots A_{N-1} R}$ and $\rho_{R}$ to get their required non-zero eigenvalues respectively. The AR $ q $-conditional entropy  $ S_q(A_1A_2 \cdots A_{N-1}|R) $ in the inertial : non-inertial bipartition always remains less than zero for both pure multiqubit GHZ and W states with one of their qubit under uniform acceleration, implying that they remain non-separable for any values of their state parameters and acceleration $\lambda$. We observed that irrespective of the value of $ q $, the non-separability of $\rho_{A_1A_2 \cdots A_{N-1} R}$ will be high at low values of acceleration and decreases with the increase in acceleration, in particular, the AR $q$-conditional entropy $ S_q(A_1A_2 \cdots A_{N-1}|R) $ tends to $ 0 $ as $ \lambda\rightarrow 1 $. The non-separability of both the pure multiqubit GHZ and W states with one of their qubits under acceleration became steeper to its approach to zero as the acceleration $\lambda$ and parameter $q$ increased. Further, we observed that the non-separability of pure multiqubit GHZ state with one of its qubits in acceleration is not dependent on the $N$, i.e., it is independent of the number of qubits in the inertial frame. However, when the number of qubits in the inertial frame of the pure multiqubit W state with one of its qubits in acceleration is increased, the non-separability decreases as $N$ increases.
 
We further extended our study to various mixed multiqubit states, $\varrho_{A_1A_2 \cdots A_{N-1} B}$, which are generated by mixing a global noise to the pure multiqubit GHZ and W states. The non-separability using the AR $q$-conditional entropy $ \mathscr{S}_q(A_1A_2 \cdots A_{N-1}|R)$  in the inertial : non-inertial bipartition is evaluated for these mixed multiqubit states, $\varrho_{A_1A_2 \cdots A_{N-1}R}$, when one of their qubits is accelerated. For these mixed multiqubit states, the AR $ q $-conditional entropy is numerically calculated by employing the {\it density matrix truncation procedure} (see Sec.\ref{subsec:dmt}). We observed that there exist regions of both separability and non-separability of the state $\varrho_{A_1A_2 \cdots A_{N-1}R}$ depending on the choices of acceleration $\lambda$, mixing parameter $ p $, and parameter $q$.  As $q$ is increased to large values, the transition curve $ \mathscr{S}_q(A_1A_2 \cdots A_{N-1}|R) =0$ becomes saturated with respect to $\lambda$ and $p$. Hence, the strongest condition on separability for these states in terms of the acceleration and mixing parameter is obtained in the asymptotic limit of $q$ tending to $\infty$. These transition points are fitted to the family of curves given by  $ (p/b)^x + \lambda^y =1 $. The separability conditions for all these mixed multiqubit GHZ and W states when one of their qubits is in uniform acceleration considered in our study are consolidated in Table \ref{Tab:separability}. For a mixed $ N $-qubit GHZ and W states with one of their qubits accelerated, we observe that the non-separable region reduces as the number of qubits in the inertial frame increases. The reduction of the non-separable region, as acceleration $\lambda$ is increased, is much faster for the $ N $-qubit mixed W state, $\varrho^{\text{W}_N}_{A_1 A_2\cdots A_{N-1}R}$, than its $ N $-qubit mixed GHZ counterpart $\varrho^{\text{GHZ}_N}_{A_1 A_2\cdots A_{N-1}R}$, indicating that, in general, GHZ states can sustain the non-separability better than W  states, when one of its qubit is uniformly accelerated with respect to others.

A comparison is made between the non-separability derived from AR $q$-conditional entropy and the entanglement measure logarithmic negativity. In single qubit accelerated $N$-qubit pure and mixed GHZ and W states, the non-separability and entanglement decrease with increasing acceleration, eventually reaching zero in the infinite acceleration limit. Moreover, the non-separability and entanglement in the single qubit accelerated pure and mixed GHZ states consistently surpass the corresponding W states for all ranges of acceleration and mixing parameters. Furthermore, in a single qubit accelerated $N$-qubit mixed GHZ and W states, we observe the transition from entanglement or non-separability to separability occur at different mixing parameter values at different accelerations. For $N = 2$, the transition curves for logarithmic negativity and AR $q$-conditional entropy are equivalent. However, as $N$ increases, AR $q$-conditional entropy imposes stricter conditions on non-separability than the logarithmic negativity.

Note that this study can be extended to fermionic systems when one of its qubits is in acceleration; however, this warrants a separate investigation. We believe that our study of non-separability in various multiqubit states with their subsystem under acceleration will pave the way for exploring numerous information-theoretic physical quantities in relativistic scenarios.  This, in turn, helps in understanding the effect of relativity on quantum systems used in space-based technology or other areas of research where relativity and quantum mechanics are considered together \cite{ursin2009space,schiller2012space}. This study also plays a pivotal role in realizing various quantum information protocols where relativistic effects cannot be neglected.

\begin{acknowledgments}
	
	We acknowledge the helpful discussions with P. Kiran. RP and HSH acknowledge financial support from the Science and Engineering Research Board (SERB), Government of India, under the CRG/2021/008795 project grant. The authors acknowledge the computations performed using the High-Performance Computing facility \textit{AnantGanak} at IIT Dharwad.
	
\end{acknowledgments}

\appendix*
\section{}
\subsection{Mixed $3 $-qubit GHZ state}\label{app:GHZ3}
A mixed $ 3 $-qubit GHZ state, with the first two qubits with Alice and the third qubit in Bob's possession, is given by
\begin{equation}\label{eq:3qubit_mixed_ghz}
	\varrho_{A_1A_2B} = \frac{p}{8} \mathbb{I}_d + (1-p) \ket{\psi_{A_1A_2B}}\bra{\psi_{A_1A_2B}},
\end{equation}
where $ \ket{\psi_{A_1A_2B}}= (1/\sqrt{2}) (\ket{000} + \ket{111}) $, $ \mathbb{I}_d $ is the $ 2^3 \times 2^3 $ identity matrix and $ p $ is the mixing parameter. Let us consider that Bob is now uniformly accelerating with respect to Alice and the resulting state, denoted as $ \varrho_{A_1A_2R_IR_{II}}^{\text{GHZ}_3} $, is obtained by using the transformations from Minkowski to Rindler spacetime in Eqs.~(\ref{eq:R0}) and (\ref{eq:R1}). Then, the mixed state $ \varrho_{A_1A_2 R}^{\text{GHZ}_3} $ after tracing out one of the disjoint regions in Rindler spacetime (region II) is given by
\begin{widetext}
	\begin{align}\label{eq:3qubit_mixedA1A2R_ghz}
		\nonumber \varrho_{A_1A_2R}^{\text{GHZ}_3} =& \frac{1}{\cosh^2 r} \sum_{n=0}^{\infty} \tanh^{2n} r  \Bigg[\frac{4-3p}{8} \left(\ket{00n}\bra{00n} + \frac{n+1}{\cosh^2 r}\ket{11n+1}\bra{11n+1} \right) \\
		&\nonumber+ \frac{1-p}{2} \frac{\sqrt{n+1}}{\cosh r} \Big(\ket{00n}\bra{11n+1} + \ket{11n+1}\bra{00n}\Big) + \frac{p}{8}\Bigg(\ket{11n}\bra{11n} + \ket{10n}\bra{10n} + \ket{01n}\bra{01n}  \\
		& + \frac{n+1}{\cosh^2 r}\Big(\ket{00n+1}\bra{00n+1} + \ket{01n+1}\bra{01n+1}  + \ket{10n+1}\bra{10n+1}\Big)\Bigg)\Bigg].
	\end{align}
\end{widetext}

\subsection{Mixed $ 3 $-qubit W state}\label{app:W3}
A mixed $ 3 $-qubit inertial state, shared between Alice and Bob, is given by
\begin{equation}\label{eq:3qubit_mixed_w}
	\varrho_{A_1A_2B}^{\text{W}_3} = \frac{p}{8} \mathbb{I}_d + (1-p) \ket{\psi_{A_1A_2B}}\bra{\psi_{A_1A_2B}}
\end{equation}
where $ \ket{\psi_{A_1A_2B}}= (1/\sqrt{3}) (\ket{001} + \ket{010} +\ket{100}) $, $ \mathbb{I}_d $ is the $ 2^3 \times 2^3 $ identity matrix and $ p $ is the mixing parameter. Let us consider that Bob is now uniformly accelerating with respect to Alice and the resulting state $ \varrho_{A_1A_2R}^{\text{W}_3} $, after tracing out the Rob's modes in region II of Rindler spacetime, is given by
\begin{widetext}
	\begin{align}\label{eq:3qubit_mixedA1A2R_w}
		&\nonumber \varrho_{A_1A_2R}^{\text{W}_3} = \frac{1}{\cosh^2 r} \sum_{n=0}^{\infty} \tanh^{2n} r  \Bigg[ \frac{8-5p}{24} \left(\ket{01n}\bra{01n} + \ket{10n}\bra{10n} + \frac{n+1}{\cosh^2 r} \ket{00n+1}\bra{00n+1} \right)   \\ &\nonumber  + \frac{p}{8}\Bigg(\ket{11n}\bra{11n} + \ket{10n}\bra{10n} +  \frac{n+1}{\cosh^2 r}\Big(\ket{00n+1}\bra{00n+1} + \ket{01n+1}\bra{01n+1} + \ket{10n+1}\bra{10n+1}\Big)\Bigg) \\
		& + \frac{1-p}{3} \left( \ket{01n}\bra{10n} + \ket{10n}\bra{01n} + \frac{\sqrt{n+1}}{\cosh r} \left( \ket{01n}\bra{00n+1} + \ket{00n+1}\bra{01n} + \ket{00n+1}\bra{10n}+\ket{10n}\bra{00n+1}\right) \right) \Bigg].	
	\end{align}
\end{widetext}

\bibliography{ref}

\begin{thebibliography}{74}%
\makeatletter
\providecommand \@ifxundefined [1]{%
 \@ifx{#1\undefined}
}%
\providecommand \@ifnum [1]{%
 \ifnum #1\expandafter \@firstoftwo
 \else \expandafter \@secondoftwo
 \fi
}%
\providecommand \@ifx [1]{%
 \ifx #1\expandafter \@firstoftwo
 \else \expandafter \@secondoftwo
 \fi
}%
\providecommand \natexlab [1]{#1}%
\providecommand \enquote  [1]{``#1''}%
\providecommand \bibnamefont  [1]{#1}%
\providecommand \bibfnamefont [1]{#1}%
\providecommand \citenamefont [1]{#1}%
\providecommand \href@noop [0]{\@secondoftwo}%
\providecommand \href [0]{\begingroup \@sanitize@url \@href}%
\providecommand \@href[1]{\@@startlink{#1}\@@href}%
\providecommand \@@href[1]{\endgroup#1\@@endlink}%
\providecommand \@sanitize@url [0]{\catcode `\\12\catcode `\$12\catcode
  `\&12\catcode `\#12\catcode `\^12\catcode `\_12\catcode `\%12\relax}%
\providecommand \@@startlink[1]{}%
\providecommand \@@endlink[0]{}%
\providecommand \url  [0]{\begingroup\@sanitize@url \@url }%
\providecommand \@url [1]{\endgroup\@href {#1}{\urlprefix }}%
\providecommand \urlprefix  [0]{URL }%
\providecommand \Eprint [0]{\href }%
\providecommand \doibase [0]{https://doi.org/}%
\providecommand \selectlanguage [0]{\@gobble}%
\providecommand \bibinfo  [0]{\@secondoftwo}%
\providecommand \bibfield  [0]{\@secondoftwo}%
\providecommand \translation [1]{[#1]}%
\providecommand \BibitemOpen [0]{}%
\providecommand \bibitemStop [0]{}%
\providecommand \bibitemNoStop [0]{.\EOS\space}%
\providecommand \EOS [0]{\spacefactor3000\relax}%
\providecommand \BibitemShut  [1]{\csname bibitem#1\endcsname}%
\let\auto@bib@innerbib\@empty
\bibitem [{\citenamefont {Bennett}\ \emph {et~al.}(1993)\citenamefont
  {Bennett}, \citenamefont {Brassard}, \citenamefont {Cr\'epeau}, \citenamefont
  {Jozsa}, \citenamefont {Peres},\ and\ \citenamefont
  {Wootters}}]{PhysRevLett.70.1895}%
  \BibitemOpen
  \bibfield  {author} {\bibinfo {author} {\bibfnamefont {C.~H.}\ \bibnamefont
  {Bennett}}, \bibinfo {author} {\bibfnamefont {G.}~\bibnamefont {Brassard}},
  \bibinfo {author} {\bibfnamefont {C.}~\bibnamefont {Cr\'epeau}}, \bibinfo
  {author} {\bibfnamefont {R.}~\bibnamefont {Jozsa}}, \bibinfo {author}
  {\bibfnamefont {A.}~\bibnamefont {Peres}},\ and\ \bibinfo {author}
  {\bibfnamefont {W.~K.}\ \bibnamefont {Wootters}},\ }\href
  {https://doi.org/10.1103/PhysRevLett.70.1895} {\bibfield  {journal} {\bibinfo
   {journal} {Phys. Rev. Lett.}\ }\textbf {\bibinfo {volume} {70}},\ \bibinfo
  {pages} {1895} (\bibinfo {year} {1993})}\BibitemShut {NoStop}%
\bibitem [{\citenamefont {Pirandola}\ \emph {et~al.}(2015)\citenamefont
  {Pirandola}, \citenamefont {Eisert}, \citenamefont {Weedbrook}, \citenamefont
  {Furusawa},\ and\ \citenamefont {Braunstein}}]{pirandola2015advances}%
  \BibitemOpen
  \bibfield  {author} {\bibinfo {author} {\bibfnamefont {S.}~\bibnamefont
  {Pirandola}}, \bibinfo {author} {\bibfnamefont {J.}~\bibnamefont {Eisert}},
  \bibinfo {author} {\bibfnamefont {C.}~\bibnamefont {Weedbrook}}, \bibinfo
  {author} {\bibfnamefont {A.}~\bibnamefont {Furusawa}},\ and\ \bibinfo
  {author} {\bibfnamefont {S.~L.}\ \bibnamefont {Braunstein}},\ }\href
  {https://www.nature.com/articles/nphoton.2015.154} {\bibfield  {journal}
  {\bibinfo  {journal} {Nature photonics}\ }\textbf {\bibinfo {volume} {9}},\
  \bibinfo {pages} {641} (\bibinfo {year} {2015})}\BibitemShut {NoStop}%
\bibitem [{\citenamefont {Hu}\ \emph {et~al.}(2023)\citenamefont {Hu},
  \citenamefont {Guo}, \citenamefont {Liu}, \citenamefont {Li},\ and\
  \citenamefont {Guo}}]{hu2023progress}%
  \BibitemOpen
  \bibfield  {author} {\bibinfo {author} {\bibfnamefont {X.-M.}\ \bibnamefont
  {Hu}}, \bibinfo {author} {\bibfnamefont {Y.}~\bibnamefont {Guo}}, \bibinfo
  {author} {\bibfnamefont {B.-H.}\ \bibnamefont {Liu}}, \bibinfo {author}
  {\bibfnamefont {C.-F.}\ \bibnamefont {Li}},\ and\ \bibinfo {author}
  {\bibfnamefont {G.-C.}\ \bibnamefont {Guo}},\ }\href
  {https://www.nature.com/articles/s42254-023-00588-x} {\bibfield  {journal}
  {\bibinfo  {journal} {Nature Reviews Physics}\ }\textbf {\bibinfo {volume}
  {5}},\ \bibinfo {pages} {339} (\bibinfo {year} {2023})}\BibitemShut {NoStop}%
\bibitem [{\citenamefont {Bennett}\ and\ \citenamefont
  {Wiesner}(1992)}]{PhysRevLett.69.2881}%
  \BibitemOpen
  \bibfield  {author} {\bibinfo {author} {\bibfnamefont {C.~H.}\ \bibnamefont
  {Bennett}}\ and\ \bibinfo {author} {\bibfnamefont {S.~J.}\ \bibnamefont
  {Wiesner}},\ }\href {https://doi.org/10.1103/PhysRevLett.69.2881} {\bibfield
  {journal} {\bibinfo  {journal} {Phys. Rev. Lett.}\ }\textbf {\bibinfo
  {volume} {69}},\ \bibinfo {pages} {2881} (\bibinfo {year}
  {1992})}\BibitemShut {NoStop}%
\bibitem [{\citenamefont {Bennett}\ and\ \citenamefont
  {Brassard}(1984)}]{bennett1984proc}%
  \BibitemOpen
  \bibfield  {author} {\bibinfo {author} {\bibfnamefont {C.}~\bibnamefont
  {Bennett}}\ and\ \bibinfo {author} {\bibfnamefont {G.~i.}\ \bibnamefont
  {Brassard}},\ }in\ \href
  {https://www.sciencedirect.com/science/article/pii/S0304397514004241?via%3Dihub#aep-article-footnote-id3}
  {\emph {\bibinfo {booktitle} {Proc. of the IEEE Int. Conf. on Computers
  Systems and Signal Processing Bangalore India}}}\ (\bibinfo {year} {1984})\
  p.\ \bibinfo {pages} {175}\BibitemShut {NoStop}%
\bibitem [{\citenamefont {Shor}\ and\ \citenamefont
  {Preskill}(2000)}]{PhysRevLett.85.441}%
  \BibitemOpen
  \bibfield  {author} {\bibinfo {author} {\bibfnamefont {P.~W.}\ \bibnamefont
  {Shor}}\ and\ \bibinfo {author} {\bibfnamefont {J.}~\bibnamefont
  {Preskill}},\ }\href {https://doi.org/10.1103/PhysRevLett.85.441} {\bibfield
  {journal} {\bibinfo  {journal} {Phys. Rev. Lett.}\ }\textbf {\bibinfo
  {volume} {85}},\ \bibinfo {pages} {441} (\bibinfo {year} {2000})}\BibitemShut
  {NoStop}%
\bibitem [{\citenamefont {Horodecki}\ \emph {et~al.}(2009)\citenamefont
  {Horodecki}, \citenamefont {Horodecki}, \citenamefont {Horodecki},\ and\
  \citenamefont {Horodecki}}]{RevModPhys.81.865}%
  \BibitemOpen
  \bibfield  {author} {\bibinfo {author} {\bibfnamefont {R.}~\bibnamefont
  {Horodecki}}, \bibinfo {author} {\bibfnamefont {P.}~\bibnamefont
  {Horodecki}}, \bibinfo {author} {\bibfnamefont {M.}~\bibnamefont
  {Horodecki}},\ and\ \bibinfo {author} {\bibfnamefont {K.}~\bibnamefont
  {Horodecki}},\ }\href {https://doi.org/10.1103/RevModPhys.81.865} {\bibfield
  {journal} {\bibinfo  {journal} {Rev. Mod. Phys.}\ }\textbf {\bibinfo {volume}
  {81}},\ \bibinfo {pages} {865} (\bibinfo {year} {2009})}\BibitemShut
  {NoStop}%
\bibitem [{\citenamefont {Modi}\ \emph {et~al.}(2012)\citenamefont {Modi},
  \citenamefont {Brodutch}, \citenamefont {Cable}, \citenamefont {Paterek},\
  and\ \citenamefont {Vedral}}]{RevModPhys.84.1655}%
  \BibitemOpen
  \bibfield  {author} {\bibinfo {author} {\bibfnamefont {K.}~\bibnamefont
  {Modi}}, \bibinfo {author} {\bibfnamefont {A.}~\bibnamefont {Brodutch}},
  \bibinfo {author} {\bibfnamefont {H.}~\bibnamefont {Cable}}, \bibinfo
  {author} {\bibfnamefont {T.}~\bibnamefont {Paterek}},\ and\ \bibinfo {author}
  {\bibfnamefont {V.}~\bibnamefont {Vedral}},\ }\href
  {https://doi.org/10.1103/RevModPhys.84.1655} {\bibfield  {journal} {\bibinfo
  {journal} {Rev. Mod. Phys.}\ }\textbf {\bibinfo {volume} {84}},\ \bibinfo
  {pages} {1655} (\bibinfo {year} {2012})}\BibitemShut {NoStop}%
\bibitem [{\citenamefont {Fuentes-Schuller}\ and\ \citenamefont
  {Mann}(2005)}]{PhysRevLett.95.120404}%
  \BibitemOpen
  \bibfield  {author} {\bibinfo {author} {\bibfnamefont {I.}~\bibnamefont
  {Fuentes-Schuller}}\ and\ \bibinfo {author} {\bibfnamefont {R.~B.}\
  \bibnamefont {Mann}},\ }\href {https://doi.org/10.1103/PhysRevLett.95.120404}
  {\bibfield  {journal} {\bibinfo  {journal} {Phys. Rev. Lett.}\ }\textbf
  {\bibinfo {volume} {95}},\ \bibinfo {pages} {120404} (\bibinfo {year}
  {2005})}\BibitemShut {NoStop}%
\bibitem [{\citenamefont {Bruschi}\ \emph {et~al.}(2010)\citenamefont
  {Bruschi}, \citenamefont {Louko}, \citenamefont {Mart\'{\i}n-Mart\'{\i}nez},
  \citenamefont {Dragan},\ and\ \citenamefont {Fuentes}}]{PhysRevA.82.042332}%
  \BibitemOpen
  \bibfield  {author} {\bibinfo {author} {\bibfnamefont {D.~E.}\ \bibnamefont
  {Bruschi}}, \bibinfo {author} {\bibfnamefont {J.}~\bibnamefont {Louko}},
  \bibinfo {author} {\bibfnamefont {E.}~\bibnamefont
  {Mart\'{\i}n-Mart\'{\i}nez}}, \bibinfo {author} {\bibfnamefont
  {A.}~\bibnamefont {Dragan}},\ and\ \bibinfo {author} {\bibfnamefont
  {I.}~\bibnamefont {Fuentes}},\ }\href
  {https://doi.org/10.1103/PhysRevA.82.042332} {\bibfield  {journal} {\bibinfo
  {journal} {Phys. Rev. A}\ }\textbf {\bibinfo {volume} {82}},\ \bibinfo
  {pages} {042332} (\bibinfo {year} {2010})}\BibitemShut {NoStop}%
\bibitem [{\citenamefont {Datta}(2009)}]{PhysRevA.80.052304}%
  \BibitemOpen
  \bibfield  {author} {\bibinfo {author} {\bibfnamefont {A.}~\bibnamefont
  {Datta}},\ }\href {https://doi.org/10.1103/PhysRevA.80.052304} {\bibfield
  {journal} {\bibinfo  {journal} {Phys. Rev. A}\ }\textbf {\bibinfo {volume}
  {80}},\ \bibinfo {pages} {052304} (\bibinfo {year} {2009})}\BibitemShut
  {NoStop}%
\bibitem [{\citenamefont {Mart\'{\i}n-Mart\'{\i}nez}\ and\ \citenamefont
  {Le\'on}(2010{\natexlab{a}})}]{PhysRevA.81.052305}%
  \BibitemOpen
  \bibfield  {author} {\bibinfo {author} {\bibfnamefont {E.}~\bibnamefont
  {Mart\'{\i}n-Mart\'{\i}nez}}\ and\ \bibinfo {author} {\bibfnamefont
  {J.}~\bibnamefont {Le\'on}},\ }\href
  {https://doi.org/10.1103/PhysRevA.81.052305} {\bibfield  {journal} {\bibinfo
  {journal} {Phys. Rev. A}\ }\textbf {\bibinfo {volume} {81}},\ \bibinfo
  {pages} {052305} (\bibinfo {year} {2010}{\natexlab{a}})}\BibitemShut
  {NoStop}%
\bibitem [{\citenamefont {Richter}\ and\ \citenamefont
  {Omar}(2015)}]{PhysRevA.92.022334}%
  \BibitemOpen
  \bibfield  {author} {\bibinfo {author} {\bibfnamefont {B.}~\bibnamefont
  {Richter}}\ and\ \bibinfo {author} {\bibfnamefont {Y.}~\bibnamefont {Omar}},\
  }\href {https://doi.org/10.1103/PhysRevA.92.022334} {\bibfield  {journal}
  {\bibinfo  {journal} {Phys. Rev. A}\ }\textbf {\bibinfo {volume} {92}},\
  \bibinfo {pages} {022334} (\bibinfo {year} {2015})}\BibitemShut {NoStop}%
\bibitem [{\citenamefont {Mart\'{\i}n-Mart\'{\i}nez}\ and\ \citenamefont
  {Le\'on}(2010{\natexlab{b}})}]{PhysRevA.81.032320}%
  \BibitemOpen
  \bibfield  {author} {\bibinfo {author} {\bibfnamefont {E.}~\bibnamefont
  {Mart\'{\i}n-Mart\'{\i}nez}}\ and\ \bibinfo {author} {\bibfnamefont
  {J.}~\bibnamefont {Le\'on}},\ }\href
  {https://doi.org/10.1103/PhysRevA.81.032320} {\bibfield  {journal} {\bibinfo
  {journal} {Phys. Rev. A}\ }\textbf {\bibinfo {volume} {81}},\ \bibinfo
  {pages} {032320} (\bibinfo {year} {2010}{\natexlab{b}})}\BibitemShut
  {NoStop}%
\bibitem [{\citenamefont {Bruschi}\ \emph {et~al.}(2012)\citenamefont
  {Bruschi}, \citenamefont {Dragan}, \citenamefont {Fuentes},\ and\
  \citenamefont {Louko}}]{PhysRevD.86.025026}%
  \BibitemOpen
  \bibfield  {author} {\bibinfo {author} {\bibfnamefont {D.~E.}\ \bibnamefont
  {Bruschi}}, \bibinfo {author} {\bibfnamefont {A.}~\bibnamefont {Dragan}},
  \bibinfo {author} {\bibfnamefont {I.}~\bibnamefont {Fuentes}},\ and\ \bibinfo
  {author} {\bibfnamefont {J.}~\bibnamefont {Louko}},\ }\href
  {https://doi.org/10.1103/PhysRevD.86.025026} {\bibfield  {journal} {\bibinfo
  {journal} {Phys. Rev. D}\ }\textbf {\bibinfo {volume} {86}},\ \bibinfo
  {pages} {025026} (\bibinfo {year} {2012})}\BibitemShut {NoStop}%
\bibitem [{\citenamefont {Buscemi}\ and\ \citenamefont
  {Bordone}(2011)}]{PhysRevA.84.022303}%
  \BibitemOpen
  \bibfield  {author} {\bibinfo {author} {\bibfnamefont {F.}~\bibnamefont
  {Buscemi}}\ and\ \bibinfo {author} {\bibfnamefont {P.}~\bibnamefont
  {Bordone}},\ }\href {https://doi.org/10.1103/PhysRevA.84.022303} {\bibfield
  {journal} {\bibinfo  {journal} {Phys. Rev. A}\ }\textbf {\bibinfo {volume}
  {84}},\ \bibinfo {pages} {022303} (\bibinfo {year} {2011})}\BibitemShut
  {NoStop}%
\bibitem [{\citenamefont {Moradi}(2009)}]{PhysRevA.79.064301}%
  \BibitemOpen
  \bibfield  {author} {\bibinfo {author} {\bibfnamefont {S.}~\bibnamefont
  {Moradi}},\ }\href {https://doi.org/10.1103/PhysRevA.79.064301} {\bibfield
  {journal} {\bibinfo  {journal} {Phys. Rev. A}\ }\textbf {\bibinfo {volume}
  {79}},\ \bibinfo {pages} {064301} (\bibinfo {year} {2009})}\BibitemShut
  {NoStop}%
\bibitem [{\citenamefont {Mart\'{\i}n-Mart\'{\i}nez}\ and\ \citenamefont
  {Fuentes}(2011)}]{PhysRevA.83.052306}%
  \BibitemOpen
  \bibfield  {author} {\bibinfo {author} {\bibfnamefont {E.}~\bibnamefont
  {Mart\'{\i}n-Mart\'{\i}nez}}\ and\ \bibinfo {author} {\bibfnamefont
  {I.}~\bibnamefont {Fuentes}},\ }\href
  {https://doi.org/10.1103/PhysRevA.83.052306} {\bibfield  {journal} {\bibinfo
  {journal} {Phys. Rev. A}\ }\textbf {\bibinfo {volume} {83}},\ \bibinfo
  {pages} {052306} (\bibinfo {year} {2011})}\BibitemShut {NoStop}%
\bibitem [{\citenamefont {Qiang}\ \emph {et~al.}(2018)\citenamefont {Qiang},
  \citenamefont {Sun}, \citenamefont {Dong},\ and\ \citenamefont
  {Dong}}]{PhysRevA.98.022320}%
  \BibitemOpen
  \bibfield  {author} {\bibinfo {author} {\bibfnamefont {W.-C.}\ \bibnamefont
  {Qiang}}, \bibinfo {author} {\bibfnamefont {G.-H.}\ \bibnamefont {Sun}},
  \bibinfo {author} {\bibfnamefont {Q.}~\bibnamefont {Dong}},\ and\ \bibinfo
  {author} {\bibfnamefont {S.-H.}\ \bibnamefont {Dong}},\ }\href
  {https://doi.org/10.1103/PhysRevA.98.022320} {\bibfield  {journal} {\bibinfo
  {journal} {Phys. Rev. A}\ }\textbf {\bibinfo {volume} {98}},\ \bibinfo
  {pages} {022320} (\bibinfo {year} {2018})}\BibitemShut {NoStop}%
\bibitem [{\citenamefont {Le\'on}\ and\ \citenamefont
  {Mart\'{\i}n-Mart\'{\i}nez}(2009)}]{PhysRevA.80.012314}%
  \BibitemOpen
  \bibfield  {author} {\bibinfo {author} {\bibfnamefont {J.}~\bibnamefont
  {Le\'on}}\ and\ \bibinfo {author} {\bibfnamefont {E.}~\bibnamefont
  {Mart\'{\i}n-Mart\'{\i}nez}},\ }\href
  {https://doi.org/10.1103/PhysRevA.80.012314} {\bibfield  {journal} {\bibinfo
  {journal} {Phys. Rev. A}\ }\textbf {\bibinfo {volume} {80}},\ \bibinfo
  {pages} {012314} (\bibinfo {year} {2009})}\BibitemShut {NoStop}%
\bibitem [{\citenamefont {Khan}\ \emph {et~al.}(2014)\citenamefont {Khan},
  \citenamefont {Khan},\ and\ \citenamefont {Khan}}]{khan2014non}%
  \BibitemOpen
  \bibfield  {author} {\bibinfo {author} {\bibfnamefont {S.}~\bibnamefont
  {Khan}}, \bibinfo {author} {\bibfnamefont {N.~A.}\ \bibnamefont {Khan}},\
  and\ \bibinfo {author} {\bibfnamefont {M.}~\bibnamefont {Khan}},\ }\href
  {https://doi.org/10.1088/0253-6102/61/3/02} {\bibfield  {journal} {\bibinfo
  {journal} {Communications in Theoretical Physics}\ }\textbf {\bibinfo
  {volume} {61}},\ \bibinfo {pages} {281} (\bibinfo {year} {2014})}\BibitemShut
  {NoStop}%
\bibitem [{\citenamefont {Ahn}\ \emph {et~al.}(2008)\citenamefont {Ahn},
  \citenamefont {Moon}, \citenamefont {Mann},\ and\ \citenamefont
  {Fuentes-Schuller}}]{ahn2008black}%
  \BibitemOpen
  \bibfield  {author} {\bibinfo {author} {\bibfnamefont {D.}~\bibnamefont
  {Ahn}}, \bibinfo {author} {\bibfnamefont {Y.}~\bibnamefont {Moon}}, \bibinfo
  {author} {\bibfnamefont {R.}~\bibnamefont {Mann}},\ and\ \bibinfo {author}
  {\bibfnamefont {I.}~\bibnamefont {Fuentes-Schuller}},\ }\href
  {https://doi.org/10.1088/1126-6708/2008/06/062} {\bibfield  {journal}
  {\bibinfo  {journal} {Journal of High Energy Physics}\ }\textbf {\bibinfo
  {volume} {2008}},\ \bibinfo {pages} {062} (\bibinfo {year}
  {2008})}\BibitemShut {NoStop}%
\bibitem [{\citenamefont {Ostapchuk}\ and\ \citenamefont
  {Mann}(2009)}]{PhysRevA.79.042333}%
  \BibitemOpen
  \bibfield  {author} {\bibinfo {author} {\bibfnamefont {D.~C.~M.}\
  \bibnamefont {Ostapchuk}}\ and\ \bibinfo {author} {\bibfnamefont {R.~B.}\
  \bibnamefont {Mann}},\ }\href {https://doi.org/10.1103/PhysRevA.79.042333}
  {\bibfield  {journal} {\bibinfo  {journal} {Phys. Rev. A}\ }\textbf {\bibinfo
  {volume} {79}},\ \bibinfo {pages} {042333} (\bibinfo {year}
  {2009})}\BibitemShut {NoStop}%
\bibitem [{\citenamefont {Alsing}\ \emph {et~al.}(2006)\citenamefont {Alsing},
  \citenamefont {Fuentes-Schuller}, \citenamefont {Mann},\ and\ \citenamefont
  {Tessier}}]{PhysRevA.74.032326}%
  \BibitemOpen
  \bibfield  {author} {\bibinfo {author} {\bibfnamefont {P.~M.}\ \bibnamefont
  {Alsing}}, \bibinfo {author} {\bibfnamefont {I.}~\bibnamefont
  {Fuentes-Schuller}}, \bibinfo {author} {\bibfnamefont {R.~B.}\ \bibnamefont
  {Mann}},\ and\ \bibinfo {author} {\bibfnamefont {T.~E.}\ \bibnamefont
  {Tessier}},\ }\href {https://doi.org/10.1103/PhysRevA.74.032326} {\bibfield
  {journal} {\bibinfo  {journal} {Phys. Rev. A}\ }\textbf {\bibinfo {volume}
  {74}},\ \bibinfo {pages} {032326} (\bibinfo {year} {2006})}\BibitemShut
  {NoStop}%
\bibitem [{\citenamefont {Wang}\ and\ \citenamefont
  {Jing}(2011)}]{PhysRevA.83.022314}%
  \BibitemOpen
  \bibfield  {author} {\bibinfo {author} {\bibfnamefont {J.}~\bibnamefont
  {Wang}}\ and\ \bibinfo {author} {\bibfnamefont {J.}~\bibnamefont {Jing}},\
  }\href {https://doi.org/10.1103/PhysRevA.83.022314} {\bibfield  {journal}
  {\bibinfo  {journal} {Phys. Rev. A}\ }\textbf {\bibinfo {volume} {83}},\
  \bibinfo {pages} {022314} (\bibinfo {year} {2011})}\BibitemShut {NoStop}%
\bibitem [{\citenamefont {Khan}(2014)}]{khan2014tripartite}%
  \BibitemOpen
  \bibfield  {author} {\bibinfo {author} {\bibfnamefont {S.}~\bibnamefont
  {Khan}},\ }\href {https://doi.org/https://doi.org/10.1016/j.aop.2014.05.022}
  {\bibfield  {journal} {\bibinfo  {journal} {Annals of Physics}\ }\textbf
  {\bibinfo {volume} {348}},\ \bibinfo {pages} {270} (\bibinfo {year}
  {2014})}\BibitemShut {NoStop}%
\bibitem [{\citenamefont {Basak}\ \emph {et~al.}(2023)\citenamefont {Basak},
  \citenamefont {Giataganas}, \citenamefont {Mondal},\ and\ \citenamefont
  {Wen}}]{PhysRevD.108.125009}%
  \BibitemOpen
  \bibfield  {author} {\bibinfo {author} {\bibfnamefont {J.~K.}\ \bibnamefont
  {Basak}}, \bibinfo {author} {\bibfnamefont {D.}~\bibnamefont {Giataganas}},
  \bibinfo {author} {\bibfnamefont {S.}~\bibnamefont {Mondal}},\ and\ \bibinfo
  {author} {\bibfnamefont {W.-Y.}\ \bibnamefont {Wen}},\ }\href
  {https://doi.org/10.1103/PhysRevD.108.125009} {\bibfield  {journal} {\bibinfo
   {journal} {Phys. Rev. D}\ }\textbf {\bibinfo {volume} {108}},\ \bibinfo
  {pages} {125009} (\bibinfo {year} {2023})}\BibitemShut {NoStop}%
\bibitem [{\citenamefont {Mart\'{\i}n-Mart\'{\i}nez}\ and\ \citenamefont
  {Le\'on}(2009)}]{PhysRevA.80.042318}%
  \BibitemOpen
  \bibfield  {author} {\bibinfo {author} {\bibfnamefont {E.}~\bibnamefont
  {Mart\'{\i}n-Mart\'{\i}nez}}\ and\ \bibinfo {author} {\bibfnamefont
  {J.}~\bibnamefont {Le\'on}},\ }\href
  {https://doi.org/10.1103/PhysRevA.80.042318} {\bibfield  {journal} {\bibinfo
  {journal} {Phys. Rev. A}\ }\textbf {\bibinfo {volume} {80}},\ \bibinfo
  {pages} {042318} (\bibinfo {year} {2009})}\BibitemShut {NoStop}%
\bibitem [{\citenamefont {Mart\'{\i}n-Mart\'{\i}nez}\ \emph
  {et~al.}(2010{\natexlab{a}})\citenamefont {Mart\'{\i}n-Mart\'{\i}nez},
  \citenamefont {Garay},\ and\ \citenamefont {Le\'on}}]{PhysRevD.82.064028}%
  \BibitemOpen
  \bibfield  {author} {\bibinfo {author} {\bibfnamefont {E.}~\bibnamefont
  {Mart\'{\i}n-Mart\'{\i}nez}}, \bibinfo {author} {\bibfnamefont {L.~J.}\
  \bibnamefont {Garay}},\ and\ \bibinfo {author} {\bibfnamefont
  {J.}~\bibnamefont {Le\'on}},\ }\href
  {https://doi.org/10.1103/PhysRevD.82.064028} {\bibfield  {journal} {\bibinfo
  {journal} {Phys. Rev. D}\ }\textbf {\bibinfo {volume} {82}},\ \bibinfo
  {pages} {064028} (\bibinfo {year} {2010}{\natexlab{a}})}\BibitemShut
  {NoStop}%
\bibitem [{\citenamefont {Ge}\ and\ \citenamefont {Kim}(2008)}]{ge2008quantum}%
  \BibitemOpen
  \bibfield  {author} {\bibinfo {author} {\bibfnamefont {X.-H.}\ \bibnamefont
  {Ge}}\ and\ \bibinfo {author} {\bibfnamefont {S.~P.}\ \bibnamefont {Kim}},\
  }\href {https://iopscience.iop.org/article/10.1088/0264-9381/25/7/075011}
  {\bibfield  {journal} {\bibinfo  {journal} {Classical and Quantum Gravity}\
  }\textbf {\bibinfo {volume} {25}},\ \bibinfo {pages} {075011} (\bibinfo
  {year} {2008})}\BibitemShut {NoStop}%
\bibitem [{\citenamefont {Pan}\ and\ \citenamefont
  {Jing}(2008)}]{PhysRevD.78.065015}%
  \BibitemOpen
  \bibfield  {author} {\bibinfo {author} {\bibfnamefont {Q.}~\bibnamefont
  {Pan}}\ and\ \bibinfo {author} {\bibfnamefont {J.}~\bibnamefont {Jing}},\
  }\href {https://doi.org/10.1103/PhysRevD.78.065015} {\bibfield  {journal}
  {\bibinfo  {journal} {Phys. Rev. D}\ }\textbf {\bibinfo {volume} {78}},\
  \bibinfo {pages} {065015} (\bibinfo {year} {2008})}\BibitemShut {NoStop}%
\bibitem [{\citenamefont {Bhattacharya}\ and\ \citenamefont
  {Joshi}(2022)}]{PhysRevD.105.065007}%
  \BibitemOpen
  \bibfield  {author} {\bibinfo {author} {\bibfnamefont {S.}~\bibnamefont
  {Bhattacharya}}\ and\ \bibinfo {author} {\bibfnamefont {N.}~\bibnamefont
  {Joshi}},\ }\href {https://doi.org/10.1103/PhysRevD.105.065007} {\bibfield
  {journal} {\bibinfo  {journal} {Phys. Rev. D}\ }\textbf {\bibinfo {volume}
  {105}},\ \bibinfo {pages} {065007} (\bibinfo {year} {2022})}\BibitemShut
  {NoStop}%
\bibitem [{\citenamefont {Mart\'{\i}n-Mart\'{\i}nez}\ \emph
  {et~al.}(2010{\natexlab{b}})\citenamefont {Mart\'{\i}n-Mart\'{\i}nez},
  \citenamefont {Garay},\ and\ \citenamefont {Le\'on}}]{PhysRevD.82.064006}%
  \BibitemOpen
  \bibfield  {author} {\bibinfo {author} {\bibfnamefont {E.}~\bibnamefont
  {Mart\'{\i}n-Mart\'{\i}nez}}, \bibinfo {author} {\bibfnamefont {L.~J.}\
  \bibnamefont {Garay}},\ and\ \bibinfo {author} {\bibfnamefont
  {J.}~\bibnamefont {Le\'on}},\ }\href
  {https://doi.org/10.1103/PhysRevD.82.064006} {\bibfield  {journal} {\bibinfo
  {journal} {Phys. Rev. D}\ }\textbf {\bibinfo {volume} {82}},\ \bibinfo
  {pages} {064006} (\bibinfo {year} {2010}{\natexlab{b}})}\BibitemShut
  {NoStop}%
\bibitem [{\citenamefont {Ball}\ \emph {et~al.}(2006)\citenamefont {Ball},
  \citenamefont {Fuentes-Schuller},\ and\ \citenamefont
  {Schuller}}]{ball2006entanglement}%
  \BibitemOpen
  \bibfield  {author} {\bibinfo {author} {\bibfnamefont {J.~L.}\ \bibnamefont
  {Ball}}, \bibinfo {author} {\bibfnamefont {I.}~\bibnamefont
  {Fuentes-Schuller}},\ and\ \bibinfo {author} {\bibfnamefont {F.~P.}\
  \bibnamefont {Schuller}},\ }\href
  {https://doi.org/https://doi.org/10.1016/j.physleta.2006.07.028} {\bibfield
  {journal} {\bibinfo  {journal} {Physics Letters A}\ }\textbf {\bibinfo
  {volume} {359}},\ \bibinfo {pages} {550} (\bibinfo {year}
  {2006})}\BibitemShut {NoStop}%
\bibitem [{\citenamefont {Fuentes}\ \emph {et~al.}(2010)\citenamefont
  {Fuentes}, \citenamefont {Mann}, \citenamefont {Mart\'{\i}n-Mart\'{\i}nez},\
  and\ \citenamefont {Moradi}}]{PhysRevD.82.045030}%
  \BibitemOpen
  \bibfield  {author} {\bibinfo {author} {\bibfnamefont {I.}~\bibnamefont
  {Fuentes}}, \bibinfo {author} {\bibfnamefont {R.~B.}\ \bibnamefont {Mann}},
  \bibinfo {author} {\bibfnamefont {E.}~\bibnamefont
  {Mart\'{\i}n-Mart\'{\i}nez}},\ and\ \bibinfo {author} {\bibfnamefont
  {S.}~\bibnamefont {Moradi}},\ }\href
  {https://doi.org/10.1103/PhysRevD.82.045030} {\bibfield  {journal} {\bibinfo
  {journal} {Phys. Rev. D}\ }\textbf {\bibinfo {volume} {82}},\ \bibinfo
  {pages} {045030} (\bibinfo {year} {2010})}\BibitemShut {NoStop}%
\bibitem [{\citenamefont {Steeg}\ and\ \citenamefont
  {Menicucci}(2009)}]{PhysRevD.79.044027}%
  \BibitemOpen
  \bibfield  {author} {\bibinfo {author} {\bibfnamefont {G.~V.}\ \bibnamefont
  {Steeg}}\ and\ \bibinfo {author} {\bibfnamefont {N.~C.}\ \bibnamefont
  {Menicucci}},\ }\href {https://doi.org/10.1103/PhysRevD.79.044027} {\bibfield
   {journal} {\bibinfo  {journal} {Phys. Rev. D}\ }\textbf {\bibinfo {volume}
  {79}},\ \bibinfo {pages} {044027} (\bibinfo {year} {2009})}\BibitemShut
  {NoStop}%
\bibitem [{\citenamefont {Wu}\ \emph {et~al.}(2022)\citenamefont {Wu},
  \citenamefont {Zeng},\ and\ \citenamefont {Liu}}]{wu2022quantum}%
  \BibitemOpen
  \bibfield  {author} {\bibinfo {author} {\bibfnamefont {S.-M.}\ \bibnamefont
  {Wu}}, \bibinfo {author} {\bibfnamefont {H.-S.}\ \bibnamefont {Zeng}},\ and\
  \bibinfo {author} {\bibfnamefont {T.}~\bibnamefont {Liu}},\ }\href
  {https://doi.org/10.1088/1361-6382/ac7508} {\bibfield  {journal} {\bibinfo
  {journal} {Classical and Quantum Gravity}\ }\textbf {\bibinfo {volume}
  {39}},\ \bibinfo {pages} {135016} (\bibinfo {year} {2022})}\BibitemShut
  {NoStop}%
\bibitem [{\citenamefont {Liu}\ \emph {et~al.}(2016)\citenamefont {Liu},
  \citenamefont {Goold}, \citenamefont {Fuentes}, \citenamefont {Vedral},
  \citenamefont {Modi},\ and\ \citenamefont {Bruschi}}]{liu2016quantum}%
  \BibitemOpen
  \bibfield  {author} {\bibinfo {author} {\bibfnamefont {N.}~\bibnamefont
  {Liu}}, \bibinfo {author} {\bibfnamefont {J.}~\bibnamefont {Goold}}, \bibinfo
  {author} {\bibfnamefont {I.}~\bibnamefont {Fuentes}}, \bibinfo {author}
  {\bibfnamefont {V.}~\bibnamefont {Vedral}}, \bibinfo {author} {\bibfnamefont
  {K.}~\bibnamefont {Modi}},\ and\ \bibinfo {author} {\bibfnamefont {D.~E.}\
  \bibnamefont {Bruschi}},\ }\href
  {https://doi.org/10.1088/0264-9381/33/3/035003} {\bibfield  {journal}
  {\bibinfo  {journal} {Classical and Quantum Gravity}\ }\textbf {\bibinfo
  {volume} {33}},\ \bibinfo {pages} {035003} (\bibinfo {year}
  {2016})}\BibitemShut {NoStop}%
\bibitem [{\citenamefont {Tian}\ \emph {et~al.}(2015)\citenamefont {Tian},
  \citenamefont {Wang}, \citenamefont {Fan},\ and\ \citenamefont
  {Jing}}]{tian2015relativistic}%
  \BibitemOpen
  \bibfield  {author} {\bibinfo {author} {\bibfnamefont {Z.}~\bibnamefont
  {Tian}}, \bibinfo {author} {\bibfnamefont {J.}~\bibnamefont {Wang}}, \bibinfo
  {author} {\bibfnamefont {H.}~\bibnamefont {Fan}},\ and\ \bibinfo {author}
  {\bibfnamefont {J.}~\bibnamefont {Jing}},\ }\href
  {https://www.nature.com/articles/srep07946} {\bibfield  {journal} {\bibinfo
  {journal} {Scientific reports}\ }\textbf {\bibinfo {volume} {5}},\ \bibinfo
  {pages} {7946} (\bibinfo {year} {2015})}\BibitemShut {NoStop}%
\bibitem [{\citenamefont {Ahmadi}\ \emph
  {et~al.}(2014{\natexlab{a}})\citenamefont {Ahmadi}, \citenamefont {Bruschi},
  \citenamefont {Sab{\'\i}n}, \citenamefont {Adesso},\ and\ \citenamefont
  {Fuentes}}]{ahmadi2014relativistic}%
  \BibitemOpen
  \bibfield  {author} {\bibinfo {author} {\bibfnamefont {M.}~\bibnamefont
  {Ahmadi}}, \bibinfo {author} {\bibfnamefont {D.~E.}\ \bibnamefont {Bruschi}},
  \bibinfo {author} {\bibfnamefont {C.}~\bibnamefont {Sab{\'\i}n}}, \bibinfo
  {author} {\bibfnamefont {G.}~\bibnamefont {Adesso}},\ and\ \bibinfo {author}
  {\bibfnamefont {I.}~\bibnamefont {Fuentes}},\ }\href
  {https://www.nature.com/articles/srep04996} {\bibfield  {journal} {\bibinfo
  {journal} {Scientific reports}\ }\textbf {\bibinfo {volume} {4}},\ \bibinfo
  {pages} {4996} (\bibinfo {year} {2014}{\natexlab{a}})}\BibitemShut {NoStop}%
\bibitem [{\citenamefont {Ahmadi}\ \emph
  {et~al.}(2014{\natexlab{b}})\citenamefont {Ahmadi}, \citenamefont {Bruschi},\
  and\ \citenamefont {Fuentes}}]{PhysRevD.89.065028}%
  \BibitemOpen
  \bibfield  {author} {\bibinfo {author} {\bibfnamefont {M.}~\bibnamefont
  {Ahmadi}}, \bibinfo {author} {\bibfnamefont {D.~E.}\ \bibnamefont
  {Bruschi}},\ and\ \bibinfo {author} {\bibfnamefont {I.}~\bibnamefont
  {Fuentes}},\ }\href {https://doi.org/10.1103/PhysRevD.89.065028} {\bibfield
  {journal} {\bibinfo  {journal} {Phys. Rev. D}\ }\textbf {\bibinfo {volume}
  {89}},\ \bibinfo {pages} {065028} (\bibinfo {year}
  {2014}{\natexlab{b}})}\BibitemShut {NoStop}%
\bibitem [{\citenamefont {Wang}\ \emph {et~al.}(2014)\citenamefont {Wang},
  \citenamefont {Tian}, \citenamefont {Jing},\ and\ \citenamefont
  {Fan}}]{wang2014quantum}%
  \BibitemOpen
  \bibfield  {author} {\bibinfo {author} {\bibfnamefont {J.}~\bibnamefont
  {Wang}}, \bibinfo {author} {\bibfnamefont {Z.}~\bibnamefont {Tian}}, \bibinfo
  {author} {\bibfnamefont {J.}~\bibnamefont {Jing}},\ and\ \bibinfo {author}
  {\bibfnamefont {H.}~\bibnamefont {Fan}},\ }\href
  {https://doi.org/https://doi.org/10.1038/srep07195} {\bibfield  {journal}
  {\bibinfo  {journal} {Scientific reports}\ }\textbf {\bibinfo {volume} {4}},\
  \bibinfo {pages} {7195} (\bibinfo {year} {2014})}\BibitemShut {NoStop}%
\bibitem [{\citenamefont {Wang}\ \emph {et~al.}(2010)\citenamefont {Wang},
  \citenamefont {Deng},\ and\ \citenamefont {Jing}}]{PhysRevA.81.052120}%
  \BibitemOpen
  \bibfield  {author} {\bibinfo {author} {\bibfnamefont {J.}~\bibnamefont
  {Wang}}, \bibinfo {author} {\bibfnamefont {J.}~\bibnamefont {Deng}},\ and\
  \bibinfo {author} {\bibfnamefont {J.}~\bibnamefont {Jing}},\ }\href
  {https://doi.org/10.1103/PhysRevA.81.052120} {\bibfield  {journal} {\bibinfo
  {journal} {Phys. Rev. A}\ }\textbf {\bibinfo {volume} {81}},\ \bibinfo
  {pages} {052120} (\bibinfo {year} {2010})}\BibitemShut {NoStop}%
\bibitem [{\citenamefont {Alsing}\ and\ \citenamefont
  {Milburn}(2003)}]{PhysRevLett.91.180404}%
  \BibitemOpen
  \bibfield  {author} {\bibinfo {author} {\bibfnamefont {P.~M.}\ \bibnamefont
  {Alsing}}\ and\ \bibinfo {author} {\bibfnamefont {G.~J.}\ \bibnamefont
  {Milburn}},\ }\href {https://doi.org/10.1103/PhysRevLett.91.180404}
  {\bibfield  {journal} {\bibinfo  {journal} {Phys. Rev. Lett.}\ }\textbf
  {\bibinfo {volume} {91}},\ \bibinfo {pages} {180404} (\bibinfo {year}
  {2003})}\BibitemShut {NoStop}%
\bibitem [{\citenamefont {Tjoa}(2022)}]{PhysRevA.106.032432}%
  \BibitemOpen
  \bibfield  {author} {\bibinfo {author} {\bibfnamefont {E.}~\bibnamefont
  {Tjoa}},\ }\href {https://doi.org/10.1103/PhysRevA.106.032432} {\bibfield
  {journal} {\bibinfo  {journal} {Phys. Rev. A}\ }\textbf {\bibinfo {volume}
  {106}},\ \bibinfo {pages} {032432} (\bibinfo {year} {2022})}\BibitemShut
  {NoStop}%
\bibitem [{\citenamefont {Grochowski}\ \emph {et~al.}(2017)\citenamefont
  {Grochowski}, \citenamefont {Rajchel}, \citenamefont {Kia\l{}ka},\ and\
  \citenamefont {Dragan}}]{PhysRevD.95.105005}%
  \BibitemOpen
  \bibfield  {author} {\bibinfo {author} {\bibfnamefont {P.~T.}\ \bibnamefont
  {Grochowski}}, \bibinfo {author} {\bibfnamefont {G.}~\bibnamefont {Rajchel}},
  \bibinfo {author} {\bibfnamefont {F.}~\bibnamefont {Kia\l{}ka}},\ and\
  \bibinfo {author} {\bibfnamefont {A.}~\bibnamefont {Dragan}},\ }\href
  {https://doi.org/10.1103/PhysRevD.95.105005} {\bibfield  {journal} {\bibinfo
  {journal} {Phys. Rev. D}\ }\textbf {\bibinfo {volume} {95}},\ \bibinfo
  {pages} {105005} (\bibinfo {year} {2017})}\BibitemShut {NoStop}%
\bibitem [{\citenamefont {Khan}\ and\ \citenamefont
  {Khan}(2015)}]{khan2015relativistic}%
  \BibitemOpen
  \bibfield  {author} {\bibinfo {author} {\bibfnamefont {S.}~\bibnamefont
  {Khan}}\ and\ \bibinfo {author} {\bibfnamefont {N.~A.}\ \bibnamefont
  {Khan}},\ }\href {https://doi.org/https://doi.org/10.1140/epjp/i2015-15216-0}
  {\bibfield  {journal} {\bibinfo  {journal} {The European Physical Journal
  Plus}\ }\textbf {\bibinfo {volume} {130}},\ \bibinfo {pages} {1} (\bibinfo
  {year} {2015})}\BibitemShut {NoStop}%
\bibitem [{\citenamefont {Haseli}(2019)}]{haseli2019quantum}%
  \BibitemOpen
  \bibfield  {author} {\bibinfo {author} {\bibfnamefont {S.}~\bibnamefont
  {Haseli}},\ }\href {https://doi.org/10.1140/epjc/s10052-019-7129-1}
  {\bibfield  {journal} {\bibinfo  {journal} {The European Physical Journal C}\
  }\textbf {\bibinfo {volume} {79}},\ \bibinfo {pages} {616} (\bibinfo {year}
  {2019})}\BibitemShut {NoStop}%
\bibitem [{\citenamefont {Khan}\ \emph {et~al.}(2022)\citenamefont {Khan},
  \citenamefont {Jan}, \citenamefont {Shah},\ and\ \citenamefont
  {Khan}}]{khan2022quantum}%
  \BibitemOpen
  \bibfield  {author} {\bibinfo {author} {\bibfnamefont {N.~A.}\ \bibnamefont
  {Khan}}, \bibinfo {author} {\bibfnamefont {M.}~\bibnamefont {Jan}}, \bibinfo
  {author} {\bibfnamefont {M.}~\bibnamefont {Shah}},\ and\ \bibinfo {author}
  {\bibfnamefont {D.}~\bibnamefont {Khan}},\ }\href
  {https://doi.org/https://doi.org/10.1016/j.aop.2022.168831} {\bibfield
  {journal} {\bibinfo  {journal} {Annals of Physics}\ }\textbf {\bibinfo
  {volume} {440}},\ \bibinfo {pages} {168831} (\bibinfo {year}
  {2022})}\BibitemShut {NoStop}%
\bibitem [{\citenamefont {G\"uhne}\ and\ \citenamefont
  {Lewenstein}(2004)}]{PhysRevA.70.022316}%
  \BibitemOpen
  \bibfield  {author} {\bibinfo {author} {\bibfnamefont {O.}~\bibnamefont
  {G\"uhne}}\ and\ \bibinfo {author} {\bibfnamefont {M.}~\bibnamefont
  {Lewenstein}},\ }\href {https://doi.org/10.1103/PhysRevA.70.022316}
  {\bibfield  {journal} {\bibinfo  {journal} {Phys. Rev. A}\ }\textbf {\bibinfo
  {volume} {70}},\ \bibinfo {pages} {022316} (\bibinfo {year}
  {2004})}\BibitemShut {NoStop}%
\bibitem [{\citenamefont {Giovannetti}(2004)}]{PhysRevA.70.012102}%
  \BibitemOpen
  \bibfield  {author} {\bibinfo {author} {\bibfnamefont {V.}~\bibnamefont
  {Giovannetti}},\ }\href {https://doi.org/10.1103/PhysRevA.70.012102}
  {\bibfield  {journal} {\bibinfo  {journal} {Phys. Rev. A}\ }\textbf {\bibinfo
  {volume} {70}},\ \bibinfo {pages} {012102} (\bibinfo {year}
  {2004})}\BibitemShut {NoStop}%
\bibitem [{\citenamefont {Cerf}\ and\ \citenamefont
  {Adami}(1997)}]{PhysRevLett.79.5194}%
  \BibitemOpen
  \bibfield  {author} {\bibinfo {author} {\bibfnamefont {N.~J.}\ \bibnamefont
  {Cerf}}\ and\ \bibinfo {author} {\bibfnamefont {C.}~\bibnamefont {Adami}},\
  }\href {https://doi.org/10.1103/PhysRevLett.79.5194} {\bibfield  {journal}
  {\bibinfo  {journal} {Phys. Rev. Lett.}\ }\textbf {\bibinfo {volume} {79}},\
  \bibinfo {pages} {5194} (\bibinfo {year} {1997})}\BibitemShut {NoStop}%
\bibitem [{\citenamefont {Horodecki}\ and\ \citenamefont
  {Horodecki}(1994)}]{Horodecki1994Quantum}%
  \BibitemOpen
  \bibfield  {author} {\bibinfo {author} {\bibfnamefont {R.}~\bibnamefont
  {Horodecki}}\ and\ \bibinfo {author} {\bibfnamefont {P.}~\bibnamefont
  {Horodecki}},\ }\href
  {https://doi.org/https://doi.org/10.1016/0375-9601(94)91275-0} {\bibfield
  {journal} {\bibinfo  {journal} {Physics Letters A}\ }\textbf {\bibinfo
  {volume} {194}},\ \bibinfo {pages} {147} (\bibinfo {year}
  {1994})}\BibitemShut {NoStop}%
\bibitem [{\citenamefont {Horodecki}\ and\ \citenamefont
  {Horodecki}(1996)}]{PhysRevA.54.1838}%
  \BibitemOpen
  \bibfield  {author} {\bibinfo {author} {\bibfnamefont {R.}~\bibnamefont
  {Horodecki}}\ and\ \bibinfo {author} {\bibfnamefont {M.}~\bibnamefont
  {Horodecki}},\ }\href {https://doi.org/10.1103/PhysRevA.54.1838} {\bibfield
  {journal} {\bibinfo  {journal} {Phys. Rev. A}\ }\textbf {\bibinfo {volume}
  {54}},\ \bibinfo {pages} {1838} (\bibinfo {year} {1996})}\BibitemShut
  {NoStop}%
\bibitem [{\citenamefont {Horodecki}\ \emph {et~al.}(1996)\citenamefont
  {Horodecki}, \citenamefont {Horodecki},\ and\ \citenamefont
  {Horodecki}}]{Horodecki1996realism}%
  \BibitemOpen
  \bibfield  {author} {\bibinfo {author} {\bibfnamefont {R.}~\bibnamefont
  {Horodecki}}, \bibinfo {author} {\bibfnamefont {P.}~\bibnamefont
  {Horodecki}},\ and\ \bibinfo {author} {\bibfnamefont {M.}~\bibnamefont
  {Horodecki}},\ }\href
  {https://doi.org/https://doi.org/10.1016/0375-9601(95)00930-2} {\bibfield
  {journal} {\bibinfo  {journal} {Physics Letters A}\ }\textbf {\bibinfo
  {volume} {210}},\ \bibinfo {pages} {377} (\bibinfo {year}
  {1996})}\BibitemShut {NoStop}%
\bibitem [{\citenamefont {Nielsen}\ and\ \citenamefont
  {Kempe}(2001)}]{PhysRevLett.86.5184}%
  \BibitemOpen
  \bibfield  {author} {\bibinfo {author} {\bibfnamefont {M.~A.}\ \bibnamefont
  {Nielsen}}\ and\ \bibinfo {author} {\bibfnamefont {J.}~\bibnamefont
  {Kempe}},\ }\href {https://doi.org/10.1103/PhysRevLett.86.5184} {\bibfield
  {journal} {\bibinfo  {journal} {Phys. Rev. Lett.}\ }\textbf {\bibinfo
  {volume} {86}},\ \bibinfo {pages} {5184} (\bibinfo {year}
  {2001})}\BibitemShut {NoStop}%
\bibitem [{\citenamefont {Tsallis}\ \emph {et~al.}(2001)\citenamefont
  {Tsallis}, \citenamefont {Lloyd},\ and\ \citenamefont
  {Baranger}}]{PhysRevA.63.042104}%
  \BibitemOpen
  \bibfield  {author} {\bibinfo {author} {\bibfnamefont {C.}~\bibnamefont
  {Tsallis}}, \bibinfo {author} {\bibfnamefont {S.}~\bibnamefont {Lloyd}},\
  and\ \bibinfo {author} {\bibfnamefont {M.}~\bibnamefont {Baranger}},\ }\href
  {https://doi.org/10.1103/PhysRevA.63.042104} {\bibfield  {journal} {\bibinfo
  {journal} {Phys. Rev. A}\ }\textbf {\bibinfo {volume} {63}},\ \bibinfo
  {pages} {042104} (\bibinfo {year} {2001})}\BibitemShut {NoStop}%
\bibitem [{\citenamefont {Abe}(2002)}]{PhysRevA.65.052323}%
  \BibitemOpen
  \bibfield  {author} {\bibinfo {author} {\bibfnamefont {S.}~\bibnamefont
  {Abe}},\ }\href {https://doi.org/10.1103/PhysRevA.65.052323} {\bibfield
  {journal} {\bibinfo  {journal} {Phys. Rev. A}\ }\textbf {\bibinfo {volume}
  {65}},\ \bibinfo {pages} {052323} (\bibinfo {year} {2002})}\BibitemShut
  {NoStop}%
\bibitem [{\citenamefont {Rossignoli}\ \emph {et~al.}(2011)\citenamefont
  {Rossignoli}, \citenamefont {Canosa},\ and\ \citenamefont
  {Ciliberti}}]{rossignoli2011generalized}%
  \BibitemOpen
  \bibfield  {author} {\bibinfo {author} {\bibfnamefont {R.}~\bibnamefont
  {Rossignoli}}, \bibinfo {author} {\bibfnamefont {N.}~\bibnamefont {Canosa}},\
  and\ \bibinfo {author} {\bibfnamefont {L.}~\bibnamefont {Ciliberti}},\ }\href
  {https://link.springer.com/article/10.1007/s10946-011-9236-9} {\bibfield
  {journal} {\bibinfo  {journal} {Journal of Russian Laser Research}\ }\textbf
  {\bibinfo {volume} {32}},\ \bibinfo {pages} {467} (\bibinfo {year}
  {2011})}\BibitemShut {NoStop}%
\bibitem [{\citenamefont {Batle}\ \emph {et~al.}(2002)\citenamefont {Batle},
  \citenamefont {Plastino}, \citenamefont {Casas},\ and\ \citenamefont
  {Plastino}}]{batle2002conditional}%
  \BibitemOpen
  \bibfield  {author} {\bibinfo {author} {\bibfnamefont {J.}~\bibnamefont
  {Batle}}, \bibinfo {author} {\bibfnamefont {A.~R.}\ \bibnamefont {Plastino}},
  \bibinfo {author} {\bibfnamefont {M.}~\bibnamefont {Casas}},\ and\ \bibinfo
  {author} {\bibfnamefont {A.}~\bibnamefont {Plastino}},\ }\href
  {https://iopscience.iop.org/article/10.1088/0305-4470/35/48/307} {\bibfield
  {journal} {\bibinfo  {journal} {Journal of Physics A: Mathematical and
  General}\ }\textbf {\bibinfo {volume} {35}},\ \bibinfo {pages} {10311}
  (\bibinfo {year} {2002})}\BibitemShut {NoStop}%
\bibitem [{\citenamefont {Batle}\ \emph {et~al.}(2003)\citenamefont {Batle},
  \citenamefont {Plastino}, \citenamefont {Casas},\ and\ \citenamefont
  {Plastino}}]{batle2003some}%
  \BibitemOpen
  \bibfield  {author} {\bibinfo {author} {\bibfnamefont {J.}~\bibnamefont
  {Batle}}, \bibinfo {author} {\bibfnamefont {A.}~\bibnamefont {Plastino}},
  \bibinfo {author} {\bibfnamefont {M.}~\bibnamefont {Casas}},\ and\ \bibinfo
  {author} {\bibfnamefont {A.}~\bibnamefont {Plastino}},\ }\href
  {https://epjb.epj.org/articles/epjb/abs/2003/19/b03297/b03297.html}
  {\bibfield  {journal} {\bibinfo  {journal} {The European Physical Journal
  B-Condensed Matter and Complex Systems}\ }\textbf {\bibinfo {volume} {35}},\
  \bibinfo {pages} {391} (\bibinfo {year} {2003})}\BibitemShut {NoStop}%
\bibitem [{\citenamefont {Tsallis}(1988)}]{tsallis1988possible}%
  \BibitemOpen
  \bibfield  {author} {\bibinfo {author} {\bibfnamefont {C.}~\bibnamefont
  {Tsallis}},\ }\href {https://link.springer.com/article/10.1007/BF01016429}
  {\bibfield  {journal} {\bibinfo  {journal} {Journal of statistical physics}\
  }\textbf {\bibinfo {volume} {52}},\ \bibinfo {pages} {479} (\bibinfo {year}
  {1988})}\BibitemShut {NoStop}%
\bibitem [{\citenamefont {Tsallis}\ \emph {et~al.}(1998)\citenamefont
  {Tsallis}, \citenamefont {Mendes},\ and\ \citenamefont
  {Plastino}}]{tsallis1998role}%
  \BibitemOpen
  \bibfield  {author} {\bibinfo {author} {\bibfnamefont {C.}~\bibnamefont
  {Tsallis}}, \bibinfo {author} {\bibfnamefont {R.}~\bibnamefont {Mendes}},\
  and\ \bibinfo {author} {\bibfnamefont {A.~R.}\ \bibnamefont {Plastino}},\
  }\href {https://www.sciencedirect.com/science/article/pii/S0378437198004373}
  {\bibfield  {journal} {\bibinfo  {journal} {Physica A: Statistical Mechanics
  and its Applications}\ }\textbf {\bibinfo {volume} {261}},\ \bibinfo {pages}
  {534} (\bibinfo {year} {1998})}\BibitemShut {NoStop}%
\bibitem [{\citenamefont {Wehrl}(1978)}]{RevModPhys.50.221}%
  \BibitemOpen
  \bibfield  {author} {\bibinfo {author} {\bibfnamefont {A.}~\bibnamefont
  {Wehrl}},\ }\href {https://doi.org/10.1103/RevModPhys.50.221} {\bibfield
  {journal} {\bibinfo  {journal} {Rev. Mod. Phys.}\ }\textbf {\bibinfo {volume}
  {50}},\ \bibinfo {pages} {221} (\bibinfo {year} {1978})}\BibitemShut
  {NoStop}%
\bibitem [{\citenamefont {Abe}\ and\ \citenamefont
  {Rajagopal}(2001)}]{abe2001nonadditive}%
  \BibitemOpen
  \bibfield  {author} {\bibinfo {author} {\bibfnamefont {S.}~\bibnamefont
  {Abe}}\ and\ \bibinfo {author} {\bibfnamefont {A.~K.}\ \bibnamefont
  {Rajagopal}},\ }\href
  {https://www.sciencedirect.com/science/article/pii/S0378437100004763?via%3Dihub}
  {\bibfield  {journal} {\bibinfo  {journal} {Physica A: Statistical Mechanics
  and its Applications}\ }\textbf {\bibinfo {volume} {289}},\ \bibinfo {pages}
  {157} (\bibinfo {year} {2001})}\BibitemShut {NoStop}%
\bibitem [{\citenamefont {Abe}\ and\ \citenamefont
  {Rajagopal}(2002)}]{abe2002towards}%
  \BibitemOpen
  \bibfield  {author} {\bibinfo {author} {\bibfnamefont {S.}~\bibnamefont
  {Abe}}\ and\ \bibinfo {author} {\bibfnamefont {A.}~\bibnamefont
  {Rajagopal}},\ }\href
  {https://www.sciencedirect.com/science/article/pii/S0960077901000467}
  {\bibfield  {journal} {\bibinfo  {journal} {Chaos, Solitons \& Fractals}\
  }\textbf {\bibinfo {volume} {13}},\ \bibinfo {pages} {431} (\bibinfo {year}
  {2002})}\BibitemShut {NoStop}%
\bibitem [{\citenamefont {Prabhu}\ \emph {et~al.}(2007)\citenamefont {Prabhu},
  \citenamefont {Usha~Devi},\ and\ \citenamefont
  {Padmanabha}}]{PhysRevA.76.042337}%
  \BibitemOpen
  \bibfield  {author} {\bibinfo {author} {\bibfnamefont {R.}~\bibnamefont
  {Prabhu}}, \bibinfo {author} {\bibfnamefont {A.~R.}\ \bibnamefont
  {Usha~Devi}},\ and\ \bibinfo {author} {\bibfnamefont {G.}~\bibnamefont
  {Padmanabha}},\ }\href {https://doi.org/10.1103/PhysRevA.76.042337}
  {\bibfield  {journal} {\bibinfo  {journal} {Phys. Rev. A}\ }\textbf {\bibinfo
  {volume} {76}},\ \bibinfo {pages} {042337} (\bibinfo {year}
  {2007})}\BibitemShut {NoStop}%
\bibitem [{\citenamefont {Vollbrecht}\ and\ \citenamefont
  {Wolf}(2002)}]{vollbrecht2002conditional}%
  \BibitemOpen
  \bibfield  {author} {\bibinfo {author} {\bibfnamefont {K.~G.~H.}\
  \bibnamefont {Vollbrecht}}\ and\ \bibinfo {author} {\bibfnamefont {M.~M.}\
  \bibnamefont {Wolf}},\ }\href
  {https://pubs.aip.org/aip/jmp/article/43/9/4299/230960/Conditional-entropies-and-their-relation-to}
  {\bibfield  {journal} {\bibinfo  {journal} {Journal of Mathematical Physics}\
  }\textbf {\bibinfo {volume} {43}},\ \bibinfo {pages} {4299} (\bibinfo {year}
  {2002})}\BibitemShut {NoStop}%
\bibitem [{\citenamefont {Sudha}\ \emph {et~al.}(2010)\citenamefont {Sudha},
  \citenamefont {Devi},\ and\ \citenamefont {Rajagopal}}]{PhysRevA.81.024303}%
  \BibitemOpen
  \bibfield  {author} {\bibinfo {author} {\bibnamefont {Sudha}}, \bibinfo
  {author} {\bibfnamefont {A.~R.~U.}\ \bibnamefont {Devi}},\ and\ \bibinfo
  {author} {\bibfnamefont {A.~K.}\ \bibnamefont {Rajagopal}},\ }\href
  {https://doi.org/10.1103/PhysRevA.81.024303} {\bibfield  {journal} {\bibinfo
  {journal} {Phys. Rev. A}\ }\textbf {\bibinfo {volume} {81}},\ \bibinfo
  {pages} {024303} (\bibinfo {year} {2010})}\BibitemShut {NoStop}%
\bibitem [{\citenamefont {Bennett}\ \emph {et~al.}(1996)\citenamefont
  {Bennett}, \citenamefont {Brassard}, \citenamefont {Popescu}, \citenamefont
  {Schumacher}, \citenamefont {Smolin},\ and\ \citenamefont
  {Wootters}}]{PhysRevLett.76.722}%
  \BibitemOpen
  \bibfield  {author} {\bibinfo {author} {\bibfnamefont {C.~H.}\ \bibnamefont
  {Bennett}}, \bibinfo {author} {\bibfnamefont {G.}~\bibnamefont {Brassard}},
  \bibinfo {author} {\bibfnamefont {S.}~\bibnamefont {Popescu}}, \bibinfo
  {author} {\bibfnamefont {B.}~\bibnamefont {Schumacher}}, \bibinfo {author}
  {\bibfnamefont {J.~A.}\ \bibnamefont {Smolin}},\ and\ \bibinfo {author}
  {\bibfnamefont {W.~K.}\ \bibnamefont {Wootters}},\ }\href
  {https://doi.org/10.1103/PhysRevLett.76.722} {\bibfield  {journal} {\bibinfo
  {journal} {Phys. Rev. Lett.}\ }\textbf {\bibinfo {volume} {76}},\ \bibinfo
  {pages} {722} (\bibinfo {year} {1996})}\BibitemShut {NoStop}%
\bibitem [{\citenamefont {Plenio}(2005)}]{plenio2005logarithmic}%
  \BibitemOpen
  \bibfield  {author} {\bibinfo {author} {\bibfnamefont {M.~B.}\ \bibnamefont
  {Plenio}},\ }\href {https://doi.org/10.1103/PhysRevLett.95.090503} {\bibfield
   {journal} {\bibinfo  {journal} {Phys. Rev. Lett.}\ }\textbf {\bibinfo
  {volume} {95}},\ \bibinfo {pages} {090503} (\bibinfo {year}
  {2005})}\BibitemShut {NoStop}%
\bibitem [{\citenamefont {Peres}(1996)}]{peres1996separability}%
  \BibitemOpen
  \bibfield  {author} {\bibinfo {author} {\bibfnamefont {A.}~\bibnamefont
  {Peres}},\ }\href {https://doi.org/10.1103/PhysRevLett.77.1413} {\bibfield
  {journal} {\bibinfo  {journal} {Phys. Rev. Lett.}\ }\textbf {\bibinfo
  {volume} {77}},\ \bibinfo {pages} {1413} (\bibinfo {year}
  {1996})}\BibitemShut {NoStop}%
\bibitem [{\citenamefont {Ursin}\ \emph {et~al.}(2009)\citenamefont {Ursin},
  \citenamefont {Jennewein}, \citenamefont {Kofler}, \citenamefont {Perdigues},
  \citenamefont {Cacciapuoti}, \citenamefont {de~Matos}, \citenamefont
  {Aspelmeyer}, \citenamefont {Valencia}, \citenamefont {Scheidl},
  \citenamefont {Acin} \emph {et~al.}}]{ursin2009space}%
  \BibitemOpen
  \bibfield  {author} {\bibinfo {author} {\bibfnamefont {R.}~\bibnamefont
  {Ursin}}, \bibinfo {author} {\bibfnamefont {T.}~\bibnamefont {Jennewein}},
  \bibinfo {author} {\bibfnamefont {J.}~\bibnamefont {Kofler}}, \bibinfo
  {author} {\bibfnamefont {J.~M.}\ \bibnamefont {Perdigues}}, \bibinfo {author}
  {\bibfnamefont {L.}~\bibnamefont {Cacciapuoti}}, \bibinfo {author}
  {\bibfnamefont {C.~J.}\ \bibnamefont {de~Matos}}, \bibinfo {author}
  {\bibfnamefont {M.}~\bibnamefont {Aspelmeyer}}, \bibinfo {author}
  {\bibfnamefont {A.}~\bibnamefont {Valencia}}, \bibinfo {author}
  {\bibfnamefont {T.}~\bibnamefont {Scheidl}}, \bibinfo {author} {\bibfnamefont
  {A.}~\bibnamefont {Acin}}, \emph {et~al.},\ }\href
  {https://www.europhysicsnews.org/articles/epn/abs/2009/03/epn20093p26/epn20093p26.html}
  {\bibfield  {journal} {\bibinfo  {journal} {Europhysics News}\ }\textbf
  {\bibinfo {volume} {40}},\ \bibinfo {pages} {26} (\bibinfo {year}
  {2009})}\BibitemShut {NoStop}%
\bibitem [{\citenamefont {Schiller}\ \emph {et~al.}(2012)\citenamefont
  {Schiller}, \citenamefont {G{\"o}rlitz}, \citenamefont {Nevsky},
  \citenamefont {Alighanbari}, \citenamefont {Vasilyev}, \citenamefont
  {Abou-Jaoudeh}, \citenamefont {Mura}, \citenamefont {Franzen}, \citenamefont
  {Sterr}, \citenamefont {Falke} \emph {et~al.}}]{schiller2012space}%
  \BibitemOpen
  \bibfield  {author} {\bibinfo {author} {\bibfnamefont {S.}~\bibnamefont
  {Schiller}}, \bibinfo {author} {\bibfnamefont {A.}~\bibnamefont
  {G{\"o}rlitz}}, \bibinfo {author} {\bibfnamefont {A.}~\bibnamefont {Nevsky}},
  \bibinfo {author} {\bibfnamefont {S.}~\bibnamefont {Alighanbari}}, \bibinfo
  {author} {\bibfnamefont {S.}~\bibnamefont {Vasilyev}}, \bibinfo {author}
  {\bibfnamefont {C.}~\bibnamefont {Abou-Jaoudeh}}, \bibinfo {author}
  {\bibfnamefont {G.}~\bibnamefont {Mura}}, \bibinfo {author} {\bibfnamefont
  {T.}~\bibnamefont {Franzen}}, \bibinfo {author} {\bibfnamefont
  {U.}~\bibnamefont {Sterr}}, \bibinfo {author} {\bibfnamefont
  {S.}~\bibnamefont {Falke}}, \emph {et~al.},\ }in\ \href
  {https://ieeexplore.ieee.org/abstract/document/6502414} {\emph {\bibinfo
  {booktitle} {2012 European Frequency and Time Forum}}}\ (\bibinfo
  {organization} {IEEE},\ \bibinfo {year} {2012})\ pp.\ \bibinfo {pages}
  {412--418}\BibitemShut {NoStop}%
\end{thebibliography}%
\bibliographystyle{apsrev4-2}
\end{document}